\documentclass[aps,prd,preprint,nofootinbib]{revtex4-2}
\usepackage[pdftex]{graphicx,xcolor,hyperref}
\usepackage{amsmath,amssymb,bm}
\usepackage{amsfonts}
\usepackage{ascmac}
\usepackage{array}
\usepackage{slashed}
\usepackage{booktabs} 
\usepackage{tikz}
\usetikzlibrary{arrows.meta, decorations.markings,decorations.pathmorphing,intersections,patterns}
\tikzset{->-/.style={
        postaction={decorate},
        decoration={markings, mark=at position .525 with {\arrow{>}}}
    },
  ->>-/.style={
    postaction={decorate},
    decoration={
      markings,
      mark=at position .475 with {\arrow{>}},
      mark=at position .575 with {\arrow{>}}
    }
  }}

\newcommand{\pinm}{\mathrm{Pin}^{-}}
\newcommand{\pinp}{\mathrm{Pin}^{+}}

\newcommand{\ZZ}{\mathbb{Z}}
\newcommand{\RPt}{\mathbb{R}\mathrm{P}^2}

\DeclareMathOperator{\sign}{\mathrm{sign}}
\DeclareMathOperator{\Pf}{\mathrm{Pf}}

\date{\today}
\begin{document}

\title{The Arf-Brown-Kervaire invariant on a lattice}
\author{Sho Araki}
\author{Hidenori Fukaya}
\author{Tetsuya Onogi}
\author{Satoshi Yamaguchi}
\affiliation{Department of physics, The University of Osaka, Toyonaka 560-0043, Japan}
\preprint{OU-HET-1294}
\begin{abstract}

We propose a lattice formulation of the Arf-Brown-Kervaire (ABK) invariant which takes values in $\mathbb{Z}_8$.
Compared to the standard $\mathbb{Z}$-valued index, the ABK invariant is more involved in that
it arises in Majorana fermion partition functions with reflection symmetry on two-dimensional non-orientable manifolds, 
and its definition contains an infinite sum over Dirac eigenvalues that requires proper regularization.
We employ the massive Wilson Dirac operator, with and without domain-walls, on standard two-dimensional square lattices, 
and use its Pfaffian for the definition. Twisted boundary conditions and cross-caps, which reverse the orientation,
are introduced to realize nontrivial topologies equipped with nontrivial $\mathrm{Pin}^{-}$ structures of Majorana fermions.
We verify numerically (and partly analytically) that our formulation on a torus, Klein bottle, 
real projective plane (as well as its triple connected sum), and two types of M\"obius strip
reproduces the known values in continuum theory.

\end{abstract}
\maketitle

\newpage
\section{introduction}
\label{sec:intro}

Topology is a key to understanding nonperturbative aspects of quantum field theories.
Due to its stableness under perturbation of the fields, deformation of the base manifolds,
renormalization group flow and so on, topology has described many nontrivial low energy
dynamics of gauge theories.
In fermionic field theories, in particular, topology of gauge fields
is closely connected via the Atiyah-Singer(AS) index theorem \cite{Atiyah:1968}
to the Dirac operator index or quantum anomalies in more general setups,
which appears in the complex phase of fermion partition functions.
Recent developments highlight that topological properties
are discussed not only on closed manifolds, but also in systems
having boundaries, where the bulk-boundary correspondence of the anomaly
plays an essential role in characterizing the symmetry-protected topological (SPT) phases
of fermions \cite{Witten:2015aba}.

In lattice gauge theory, which is another powerful nonperturbative tool for analyzing quantum field theories,
topology is apparently a difficult property to formulate due to the absence of continuity.
In fact, it was shown in \cite{Neuberger:1997fp,Hasenfratz:1998ri} that the index
can be formulated in a similar way to the continuum theory using lattice Dirac operators
satisfying the Ginsparg-Wilson relation \cite{Ginsparg:1981bj}.
The index was defined by the Dirac operator zero modes, thanks to
the exact chiral symmetry on a lattice \cite{Luscher:1998pqa}.
In recent studies \cite{Clancy:2023ino}
the Ginsparg-Wilson relation was generalized
to wider fermionic systems such as those in odd dimensions, those
with discrete symmetries/anomalies, and those having real structures, 
to describe more nontrivial indices of Dirac operators.
However, these Dirac indices defined by the zero modes
are formulated essentially on a flat periodic square lattice,
whose continuum counterpart is limited to a flat torus.
In particular, the Ginsparg-Wilson relation is known to be broken
when we impose nontrivial boundary conditions \cite{Luscher:1998pqa}.

In \cite{Aoki:2024sjc,Aoki:2025gca}, in which three of the authors are involved,
a new generalization was proposed using K-theory \cite{MR1043170,Karoubi}.
By the so-called suspension isomorphism of K groups,
the Dirac operator index is known to be expressed by a certain type of
the spectral flow of massive Dirac operators in continuum theory.
In \cite{Fukaya:2017kpy,Fukaya:2019qlf,Fukaya:2020tjk},
it was shown that the Atiyah-Patodi-Singer(APS) index \cite{APS}
on a manifold with boundary
is also mathematically equal to the spectral flow of Dirac operators
with a domain-wall in the mass term.
The new formulation of the index was given by the
lattice version of the corresponding spectral flow.
Since the continuum target is a massive Dirac operator,
we do not need the Ginsparg-Wilson relation or any other sophisticated symmetry on a lattice,
and the Wilson Dirac operator is good enough for the formulation.
The new formulation of the index measured by the spectral flow of the Wilson Dirac operator
covers the standard AS index, APS index whose boundary may be curved,
and their mod-two versions when the Dirac operator has a real structure.
In Ref.~\cite{Aoki:2025gca} it was numerically shown that
the both $\ZZ$ and $\ZZ_2$ indices
on a torus, a stripe with two flat boundaries, a disk, and a torus with a hole,
can be evaluated on a lattice.

In this work, we try to formulate a further nontrivial topological quantity by fermion path integral on a lattice.
In continuum theory, there exist more exotic $\ZZ_8$ or $\ZZ_{16}$-valued topological invariants\footnote{
In previous studies, a state sum combinatorial definition of the ABK
invariant on lattice using Gu-Wen Grassman integral \cite{Gu:2012ib}
was formulated given by \cite{Kobayashi:2019xxg,Inamura:2019hpu},
which corresponds to regularizing topological quantum field theories
on the lattice. We also refer the readers to
\cite{Shapourian:2016kvr,Shiozaki:2017ive} in which the ABK invariant
was discussed using the Hamiltonian formalism of fermions. In this
work, we employ Euclidean fermion path-integral approach.
} \cite{Kapustin:2014dxa}.
In physics, these invariants are related to the 
topological phases of Majorana fermions protected by the time-reversal symmetry.
We will show that the $\ZZ_8$-valued Arf-Brown-Kervaire (ABK) invariant \cite{Brown}
on a two-dimensional manifold can be described by the lattice Wilson Dirac operator.
Specifically it appears in the 
phase factor of the fermion partition function $Z$ \cite{Witten:2015aba}\cite{Kaidi:2019tyf}:
\begin{align}
  Z \propto \exp\left(\frac{i\pi}{4}\beta\right), \label{eq:Z8phase}
\end{align}
where the integer-valued invariant $\beta$ modulo 8 is the ABK invariant.
Our goal is to regularize $Z$ with the Wilson Dirac operator on a lattice
and show that the ABK invariant is obtained from the complex phase of $Z$.
This requires a quite nontrivial setup on a lattice as listed below.

In continuum theory, the ABK invariant appears in the partition function 
of Majorana fermions having a reflection symmetry with $R^2=(-1)^F$ where $(-1)^F$ is the fermion number parity.
This indicates that we must have a real structure in the corresponding lattice fermion Dirac operator
with the $R$ symmetry 
and consider the Pfaffian of it, rather than its determinant.

The ABK invariant $\beta$ can be viewed as an extension of the mod-two index. In contrast to the mod-two index, which is defined only for orientable manifolds, the ABK invariant is defined even on non-orientable manifolds.
The reflection symmetry makes it possible to define the Majorana fermion on such manifolds.
Mathematically, this corresponds to introducing a $\pinm$ structure (minus sign comes from $R^2=(-1)^F$) on the manifolds.
For this requirement, we must impose ``twisted'' boundary conditions \cite{Mages:2015scv}\cite{Seiberg:2026icc} which reverse the orientation,
rather than the standard periodic or antiperiodic lattice boundary conditions.
We also note the recent related study \cite{Seiberg:2026icc} of staggered fermions on the torus and the Klein bottle, where the mod-eight anomaly (a similar mod-eight structure arises in our construction as an anomaly of the boundary fermions) were analyzed through the matching between lattice and continuum symmetries.

It is impossible to describe the ABK invariant with the zero modes of the Dirac operator alone.
It is described in terms of $\eta$-invariant \cite{APS} (spectral asymmetry between positive and negative eigenvalues) of the Dirac operator. 
The $\ZZ_8$ quantization arises from the subtle spectral asymmetry \cite{Debray:2018wfz,Zhang1994,Zhang2017}, 
and a proper treatment of regularization is required.
It is almost obvious from this fact that the overlap fermion having exact chiral zero modes
would not be useful that much.
On our two-dimensional lattices, it is manageable to numerically evaluate
the Wilson fermion Pfaffians directly and examine if the 
lattice Wilson operator is adequate for the regularization of the ABK invariant.

Finally, the ABK invariant on a manifold with boundaries
requires the APS boundary condition in the standard mathematical setup,
which is nonlocal and difficult to realize on a lattice.
In our formulation, we introduce the domain-wall mass term \cite{Kaplan:1992bt},
which compensates the APS boundary condition, 
following the discussion in \cite{Fukaya:2017kpy,Fukaya:2019qlf,Fukaya:2020tjk}
and related argument in \cite{Witten:2019bou}.
We note that if the lattice version of this formulation is successful, 
it offers a nonperturbative path-integral regularization of the one-dimensional Majorana chain
discussed in \cite{Fidkowski:2009dba,Fidkowski:2010jmn}
and the anomaly inflow between its bulk and edge.

In this work, we evaluate the Majorana fermion Pfaffian 
employing the Wilson Dirac operator on various types of two-dimensional lattices
with and without domain-walls.
We find that the extracted ABK invariant is quantized into $\ZZ_8$ values
on a conventional flat torus, 
klein bottle, real projective plane as well as its triple connected sum, 
and two different types of the M\"obius strip. 
In the torus case, reproducing the continuum ABK invariant (i.e., the mod-two index, which is the special case for orientable manifolds) can be checked rather easily.
In contrast, the corresponding statement on non-orientable manifolds is nontrivial.
Performing the analysis on several lattices with different lattice spacings,
we examine their convergence to the continuum values.
For the Klein bottle case, we successfully decompose the Pfaffian into Fourier modes and identify a specific set of modes that contributes to the ABK invariant.

We expect that the analogous $\ZZ_{16}$-valued invariant\footnote{A state sum combinatorial definition of this invariant on lattice using Gu-Wen Grassman integral \cite{Gu:2012ib} is given by \cite{Kobayashi:2019xxg,Tata:2021jwp}} in higher dimensions
can be defined in a similar way, although there may be technical difficulties due to numerical costs.

The rest of this paper is organized as follows. In Sec.~\ref{sec:continuum}, 
we review how the ABK invariants arise in the continuum theory of Majorana fermions. 
In Sec.~\ref{ss:lattice_setup}, we turn to the lattice path integral formulation
and introduce various unorientable manifolds with various $\pinm$ structures by twisted boundary conditions
as well as cross-caps with and without domain-walls.
In Sec.~\ref{sec:Fourier}, for the cases of a torus and Klein bottle,
we analytically compute the ABK invariant on the corresponding lattices.
Then we give our numerical results for more general manifolds in Sec.~\ref{ss:numerical_calculation}.
In Sec.~\ref{sec:summary}, we give a summary and discussions.

\section{Two-dimensional Majorana fermion and the ABK invariant}
\label{sec:continuum}

We first consider in continuum theory the Majorana fermion partition functions on a two-dimensional manifold $X$
and introduce the ABK invariant.

Majorana fermions are obtained by imposing on the standard Dirac fermion $\psi$,
the Majorana condition $\bar{\psi}=\psi^T C$ with the charge conjugation matrix $C$.
For the Dirac matrices satisfying $\left\{ \gamma^a, \gamma^b \right\} = \delta^{ab}$,
$C$ operates as $C\gamma^a C^{-1} = -(\gamma^a)^T$.
Our action to be considered is
\begin{equation}
  S =   \int_{X}^{} d^2x \sqrt{g}  \psi^T C (D + m) \psi, \label{eq:continuum_action}
\end{equation}
where $D=\gamma^a e_a^\mu D_\mu$ is the Dirac operator with the zweibein $e_a^\mu$, which is related to the metric through
$g^{\mu\nu}=e_a^\mu e_b^\nu\delta^{ab}$, $g=\det g_{\mu\nu}$ and $D_\mu=\partial_\mu + \omega_\mu$ is the covariant derivative
with the spin connection $\omega_\mu$.

We take the standard representation of the Dirac matrices by the Pauli matrices:
\begin{align}
  \gamma^1 = \begin{pmatrix} 0 & 1 \\ 1 & 0  \end{pmatrix}, && \gamma^2 = \begin{pmatrix} 0 & -i \\ i & 0  \end{pmatrix}, \label{eq:gamma_def}
\end{align}
and 
\begin{equation}
  C = \begin{pmatrix} 0 & -1 \\ 1 & 0  \end{pmatrix}. \label{eq:Cmat_def}
\end{equation}
Note that $C$ is anti-symmetric, so that the mass term $m \psi^T C \psi$ is generally nonzero.
Here we also define the chirality matrix $\bar{\gamma} = -i\gamma^1 \gamma^2={\rm diag}(1,-1)$.

In order to make the manifold $X$ endowed with a $\pinm$ structure, 
we require the fermion action to have a reflection symmetry.

Here we take the $x$ direction 
to be reversed under the reflection and the fermion fields are transformed as
\begin{equation}
  \psi(x,y) \to R_x \psi(x,y) = \gamma^a e_a^1 \bar{\gamma} \psi(-x,y),
\end{equation}
as well as the metric and zweibein accordingly.
Note that the spin connection $\omega_\mu$ transforms like an axial vector, since
it is proportional to $[\gamma^1,\gamma^2]/2=i\bar{\gamma}$.
 
Twice of this transformation yields $R_x^2=-1$, or $R_x^2=(-1)^F$ with the fermion number $F$ to be precise,
in contrast to the $\pinp$ case (where the reflections twice give identity),
the sign of the mass term $m\psi^T C \psi$ is unchanged.
The chirality is no more good quantum number in the theory even when $m=0$,
since $\bar{\gamma}$ does not commute with $R_x$.
Using this $R_x$ symmetry together with appropriate boundary conditions,
we can consider various unorientable manifolds with various $\pinm$ structures
 and the Majorana fermions on them\footnote{
The partition function on non-orientable manifolds in the operator formalism was discussed in \cite{Shapourian:2016kvr,Shiozaki:2017ive}.
}.

The partition function of Majorana fermions is formally given by the Pfaffian
of the operator $C(D+m)$ which needs to be properly regularized.
Here we employ the Pauli-Villars regularization
with a bosonic spinor with the mass value $M$.
Note that for every eigenfunction $\chi$ with a pure imaginary eigenvalue
$i\lambda$ of $D$,  $C\chi^*$ is another eigenmode with the same eigenvalue
$i\lambda$.
Due to this property, the regularized Majorana fermion partition function
can be expressed by
\begin{equation}
  \label{eq:pfaffian-eta}
Z=\frac{\mathrm{Pf}[C(D+m)]}{\mathrm{Pf}[C(D+M)]} = {\prod_\lambda^{}}'\frac{i\lambda +m}{i\lambda+M},
\end{equation}
where $\prod'$ means that the product is taken over only one eigenvalue from
the doubly degenerate pair.

The relative sign of the fermion mass and the Pauli-Villars mass
is important in that the ABK invariant appears only when $mM<0$,
where the system is in a nontrivial topological phase.
Following \cite{Witten:2015aba}, let us take the  following limit,
 $-m=M\to +\infty$ to obtain
\begin{equation}
 Z 
  = \lim_{M\to +\infty}{\prod_{\lambda}^{}}'\frac{\lambda +iM }{\lambda} \frac{\lambda}{\lambda -iM} .
\label{eq:Z_decomposition}
\end{equation}
Intuitively, for large $M$, the phase of this infinite product can be rewritten as,  
\begin{equation}
  \lim_{M\to +\infty}{\prod_{\lambda}^{}}'\frac{\lambda +iM }{\lambda} \frac{\lambda}{\lambda -iM} = \lim_{M\to +\infty}{\prod_{\lambda}^{}}'\left(\frac{iM }{\lambda}\right)\left( \frac{-\lambda}{iM}\right)\propto 
  \prod_{\lambda}^{}\left( -1 \right)\label{eq:prod_sign}
\end{equation}
If the spectrum is paired as $\pm i\lambda$ (which happens when the spacetime manifold is orientable
\footnote{On orientable manifolds, one can define a globally well-defined chirality operator that commutes with the free Dirac operator $D$, implying that the spectrum is paired as $\pm i\lambda$. For non-orientable manifolds, no globally well-defined chirality operator exists, and the $\pm$-pairing of the spectrum is lost.}), the infinite product reduces to the number of unpaired $\lambda=0$ modes of the Dirac operators.
Dividing this number by 2 (since the eigenvalues are always doubly degenerate) gives the mod-two index $I(D)$ of $D$, and the phase of the partition function can then be written as
\begin{equation}
  Z\propto  (-1)^{I(D)} .
\end{equation} 
However, in general situation, the eigenvalues are not paired as $\pm i\lambda$. Then the infinite product of the right hand side of Eq.\eqref{eq:prod_sign} is not well-defined. Instead we rewrite this phase in terms of an  infinite sum over $\lambda$,
\begin{equation}
  \exp\left[i\frac{\pi}{2} \left( \sum_{\lambda\geq0}1 - \sum_{\lambda<0}1\right) \right].\label{eq:non_regEta}
\end{equation}
From the left hand side of Eq.\eqref{eq:prod_sign}, one can see that the complex phase of $\frac{\lambda +iM }{\lambda} \frac{\lambda}{\lambda -iM}$ is mapped to slightly different regions depending on $\sign(\lambda)$.
Namely,
\begin{equation}
  0<\arg\left( \frac{\lambda +iM }{\lambda -iM} \right)\le\pi\quad(\lambda\ge 0),\qquad -\pi<\arg\left( \frac{\lambda +iM }{\lambda -iM} \right)<0\quad(\lambda\le 0).
\end{equation} Therefore, this infinite sum can be regarded as accumulating the phase contributions according to the sign of $\lambda$.
An appropriate regularization of this infinite sum yields the $\eta$ invariant \cite{APS} \cite{Witten:2015aba}, which provides the total phase,
\begin{equation}
 Z  \propto \exp\left[i\frac{\pi}{2} \eta(iD) \right].
\end{equation}
The $\eta$ invariant is usually defined by the zeta-function regularization as:
\begin{equation}
  \eta(iD) = \lim_{s \to +0 }\sum_{\lambda}^{} \sign (\lambda) \frac{1}{|\lambda|^s},
\end{equation}
with the $\sign$ function defined by
\begin{equation}
  \sign(\lambda) =\begin{cases}
    +1 & \text{if $\lambda \geq 0$}\\
    -1 & \text{if $\lambda < 0$}
  \end{cases}.
\end{equation}
Here the summation is taken over all eigenvalues $\lambda$ of the Dirac operator.
Note that the exponent in Eq.~(\ref{eq:pfaffian-eta})
has $O(1/M)$ corrections when $M$ is finite.

According to Refs.\cite{Debray:2018wfz,Zhang1994,Zhang2017},
the ABK invariant\footnote{
The original definition of the ABK invariant \cite{Brown} was given in terms of $\ZZ_2$-cohomology of manifolds, 
and it is a topological invariant depending only on global topology of manifold and its $\pinm$ structure, 
independent of the metric $g$. Furthermore, a correspondence with the two-dimensional $\pinm$ bordism group $\Omega_2^{\pinm}\simeq \ZZ_8$ is known \cite{Kirby_Taylor_1991}. 
Indeed, the two-dimensional real projective plane ($\RPt$) which generates the bordism group, 
takes the minimal value $\beta=\pm 1$ (where the sign depends on the choice of $\pinm$ structure).
} multiplied by 1/2 can be identified as a special version
of the $\eta$ invariant, which takes only integral or half-integral values
on $\pinm$ manifolds.
Namely, we can denote it by
\begin{equation}
 \beta=2\eta(iD)\;\;\; \mbox{mod 8}.
\end{equation}
We also note that $\eta$ takes even integers on orientable manifolds
so that the Pfaffian is always real and the invariant reduces to $\ZZ_2$ values.
Therefore, we need to consider unorientable manifolds to obtain nontrivial $\ZZ_8$ values of the ABK invariant.

For the Pfaffian of fermions in a two-dimensional 
condensed matter, the $\ZZ_8$ quantization classifies the SPT phases.
When the number of fermion flavors is $k=1,\dots,7 \mod 8$, the complex phase $\exp\left(i\frac{\pi}{4}k\beta\right)$  for $m<0$ cannot be modified to 
unity in the case of $m>0$ without hitting singularity (or gap closing).
In the $k=0 \mod 8$ case, however, there is no relative phase difference and there is no topological obstruction. 
As Fidkowski and Kitaev \cite{Fidkowski:2009dba,Fidkowski:2010jmn} explicitly showed, 
the two partition functions ($m<0$ and $m>0$) are smoothly connected via additional four-fermi interaction terms. 
Such an argument requires access to manifolds with odd $\beta$. If one looks only at orientable examples like the torus (with $\beta=0,4$), one can at best detect whether $k$ is even or odd\footnote{This is related to $\ZZ$ classification of SPT phases without the reflection symmetry.}.

So far we have implicitly assumed that $X$ is a closed manifold.
Now let us discuss the case with boundaries.
In mathematics, the $\eta$ invariant of the massless Dirac operator $iD$ can be
defined on such a manifold with
the so-called APS boundary condition \cite{APS}.
As described in Ref.\cite{Witten:2015aba}, this $\eta$ invariant
with the APS boundary condition describes the anomaly inflow 
or anomaly cancellation between the bulk and edge fermions in the SPT phases.
However, the APS condition is nonlocal and unlikely to be 
realized in physics.

The massive fermion system follows a more natural boundary 
condition which is different from the APS condition.
It is given by the projection with respect to the Dirac matrix $\gamma_\perp$
which is perpendicular to the surface.
This boundary condition  is the same as that of 
the domain-wall fermion \cite{Jackiw:1975fn,Callan:1984sa}
where the mass term flips its sign at the wall,
on which the massless chiral modes are localized.

In Refs.\cite{Fukaya:2017kpy,Fukaya:2019qlf}, 
it was shown that the APS index \cite{APS}
on a manifold with boundary is mathematically equal to 
the $\eta$ invariant of the domain-wall fermion Dirac operator,
where the original boundary is extended to form a closed manifold
and the domain-wall is instead put at the original location of 
the boundary. 
Interestingly, this expression of the index does not depend on 
the extended subregion at all.
This physicist-friendly expression of the APS index 
is formulated in the lattice gauge theory \cite{Aoki:2024sjc,Aoki:2025gca}, too. 
With a different discussion \cite{Witten:2019bou}, more general $\eta$ invariant
of the massless Dirac operator was shown to be equal to 
the phase of the massive fermion partition function with the physical boundary condition.

In this work, we examine if the phase of the 
lattice domain-wall fermion partition function \cite{Kaplan:1992bt,Shamir:1993zy} is consistent with the $\ZZ_8$-valued
$\eta$ invariant in the corresponding continuum theory, namely, that of massless Dirac operator with the APS boundary condition.
This is a nontrivial equality since it is impossible
to define a massless Dirac operator on a lattice 
without breaking the Ginsparg-Wilson relation
nor any boundary condition which corresponds to the APS boundary condition.
The spectrum of the domain-wall Dirac operator 
is totally different from that of the target 
massless Dirac operator in the continuum theory.

\section{A Lattice regularization of the ABK invariant} \label{ss:lattice_setup}

We consider standard two-dimensional square lattices whose size is denoted by $N_x a\times N_y a$
with the lattice spacing $a$.
On each of these lattices
we discretize the action Eq.~\eqref{eq:continuum_action} by that of the Wilson fermion
\begin{equation}
  S_{\text{W}} = \sum_{x,y} a^2 \frac{i}{2} \psi(x,y)^T C D_W(m) \psi(x,y),
\end{equation}
where $x$ and $y$ take integer multiple of $a$: $0,a,2a\cdots (N_{x/y}-1)a$, 
and  $D_W(m)$ denotes the massive Wilson-Dirac operator
\begin{equation}
  D_W(m)=
\sum_\mu \gamma^\mu \frac{\nabla^f_\mu + \nabla^b_\mu}{2}+ m + \sum_\mu r\frac{\nabla^f_\mu - \nabla^b_\mu}{2},
\end{equation}
where $\nabla^{f}_\mu$ and $\nabla^{b}_\mu$ are the forward and backward difference operators, respectively.
The definitions of $\gamma^\mu$'s and $C$ matrix are the same as those given in 
Eqs.~\eqref{eq:gamma_def} and \eqref{eq:Cmat_def}.
Here we take a constant mass term $m$ but later it will be generalized to the case with domain-walls.
Since the Majorana fermions we consider in this work are neutral to any gauge interactions, 
we take all the link variables unity, except for those around singularities explained below.

In the following, the Wilson parameter is set to $r=1$.
We have implicitly assumed that the zweibein in our target continuum theory is 
aligned with the coordinate ($e^\mu_a =\delta^\mu_a$).

The partition function of the lattice action is given by the Pfaffian
\begin{equation}
  Z(m)=\mathrm{Pf}\left( C D_W(m) \right).
\end{equation}
For an antisymmetric matrix A of finite size $2n\times 2n$, its Pfaffian is defined as 
\begin{equation}
  \Pf(A) = \frac{1}{2^n n!} \sum_{\sigma \in S_{2n}} \sign(\sigma) A_{\sigma(1)\sigma(2)}A_{\sigma(3)\sigma(4)}\cdots A_{\sigma(2n-1)\sigma(2n)},
\end{equation}
where $S_{2n}$ denotes the set of permutations of $\{1,2,\cdots,2n\}$.
The Pfaffian is also known as the square root of the determinant, $(\Pf A)^2 = \det A$, yet the expression above defines it as a polynomial in the matrix elements with no sign ambiguity.
Numerically, $\mathrm{Pf}\left( C D_W(m) \right)$ can be  evaluated with a reasonable choice of 
the sizes $N_x, N_y< 30$. 
By imposing different boundary conditions, the operator $C D_W(m)$ becomes 
a finite-sized complex matrix so that $Z(m)$ can have a nontrivial phase,
which defines our lattice formulation of the ABK invariant:
\begin{equation}
 \beta^{\mathrm{lat}}(m,a):= \frac{4}{\pi i}\arg \left( \frac{\mathrm{Pf}\left( C D_{W}(m) \right)}{\mathrm{Pf}\left( C D_{W}(|m|)\right)} \right).
\end{equation}
Here, the Pauli-Villar's partition function $\mathrm{Pf}\left( C D_{W}(|m|)\right)=Z(|m|)$ is introduced 
to compare the results with the continuum theory\footnote{
It is expected that the phase of $\mathrm{Pf}\left( C D_{W}(|m|)\right)$ is trivial, at least, in the continuum limit.
}.

Our goal of this work is to examine 
if $\beta^{\mathrm{lat}}(m, a)$ is consistent with the continuum value
or not in the $a\to 0$ and $-m\to \infty$ limits.
Our approach on the standard square lattice to set up non-trivial 
$\pinm$ manifolds by the boundary conditions 
is distinct from those employing triangular lattices \cite{Brower:2016vsl}
to express nonzero curvatures.

\subsection{Flat torus $T^2$}
\begin{figure}[htbp]
    \centering
    \begin{tikzpicture}[scale=1.2, line width=1pt
]

    \node at (1,-0.7) {(a) Torus};
    \draw (0,0) rectangle (2,2);

    \draw[->-,blue] (0,0) -- (0,2);
    \draw[->-,blue] (2,0) -- (2,2);

    \draw[->>-,red] (0,0) -- (2,0);
    \draw[->>-,red] (0,2) -- (2,2);
    \def\N{4}          
    \def\a{0.5}        
    \def\LDx{0.25}
    \def\LDy{0.25}
    \def\margin{0.4}  
    \def\latticeColor{gray!50}
        
    \foreach \j in {0,...,\numexpr\N-1\relax}{
      \draw[\latticeColor] (-\margin+\LDx,\j*\a+\LDy)--(\N*\a-1*\a+\margin+\LDx,\j*\a+\LDy);
    }
        
    \foreach \i in {0,...,\numexpr\N-1\relax}{
      \draw[\latticeColor] (\i*\a+\LDx, -\margin+\LDy) -- (\i*\a+\LDx, \a*\N-\a + \margin+\LDy);
    }
    \node[blue!70] at (-0.3,0.25) {0};
    \node[blue!70] at (-0.3,0.75) {1};
    \node[blue!70] at (-0.3,1.25) {2};
    \node[blue!70] at (-0.3,1.75) {3};
    \node[blue!70] at (2.3,0.25) {0};
    \node[blue!70] at (2.3,0.75) {1};
    \node[blue!70] at (2.3,1.25) {2};
    \node[blue!70] at (2.3,1.75) {3};
    \node[red!70] at (0.25,-0.3) {0};
    \node[red!70] at (0.75,-0.3) {1};
    \node[red!70] at (1.25,-0.3) {2};
    \node[red!70] at (1.75,-0.3) {3};
    \node[red!70] at (0.25,2.3) {0};
    \node[red!70] at (0.75,2.3) {1};
    \node[red!70] at (1.25,2.3) {2};
    \node[red!70] at (1.75,2.3) {3};
    \begin{scope}[xshift=4.0cm]
        \node at (1,-0.7) {(b) Klein Bottle};
        \draw (0,0) rectangle (2,2);

        \draw[->-,blue] (0,0) -- (0,2);
        \draw[->-,blue] (2,0) -- (2,2);

        \draw[->>-,red] (0,0) -- (2,0);
        \draw[->>-,red] (2,2) -- (0,2);    
        \def\N{4}          
        \def\a{0.5}        
        \def\LDx{0.25}
        \def\LDy{0.25}
        \def\margin{0.4}  
        \def\latticeColor{gray!50}
            
        \foreach \j in {0,...,\numexpr\N-1\relax}{
          \draw[\latticeColor] (-\margin+\LDx,\j*\a+\LDy)--(\N*\a-1*\a+\margin+\LDx,\j*\a+\LDy);
        }
            
        \foreach \i in {0,...,\numexpr\N-1\relax}{
          \draw[\latticeColor] (\i*\a+\LDx, -\margin+\LDy) -- (\i*\a+\LDx, \a*\N-\a + \margin+\LDy);
        }
        \node[blue!70] at (-0.3,0.25) {0};
        \node[blue!70] at (-0.3,0.75) {1};
        \node[blue!70] at (-0.3,1.25) {2};
        \node[blue!70] at (-0.3,1.75) {3};
        \node[blue!70] at (2.3,0.25) {0};
        \node[blue!70] at (2.3,0.75) {1};
        \node[blue!70] at (2.3,1.25) {2};
        \node[blue!70] at (2.3,1.75) {3};
        \node[red!70] at (0.25,-0.3) {0};
        \node[red!70] at (0.75,-0.3) {1};
        \node[red!70] at (1.25,-0.3) {2};
        \node[red!70] at (1.75,-0.3) {3};
        \node[red!70] at (0.25,2.3) {3};
        \node[red!70] at (0.75,2.3) {2};
        \node[red!70] at (1.25,2.3) {1};
        \node[red!70] at (1.75,2.3) {0};
    \end{scope}

    \begin{scope}[xshift=8.0cm]
        \node at (1,-0.7) {(c) Real projective plane ($\RPt$)};
        \draw (0,0) rectangle (2,2);
        
        \draw[->-,blue] (0,0) -- (0,2);
        \draw[->-,blue] (2,2) -- (2,0);

        \draw[->>-,red] (0,0) -- (2,0);
        \draw[->>-,red] (2,2) -- (0,2);    
        \def\N{4}          
        \def\a{0.5}        
        \def\LDx{0.25}
        \def\LDy{0.25}
        \def\margin{0.4}  
        \def\latticeColor{gray!50}
            
        \foreach \j in {0,...,\numexpr\N-1\relax}{
          \draw[\latticeColor] (-\margin+\LDx,\j*\a+\LDy)--(\N*\a-1*\a+\margin+\LDx,\j*\a+\LDy);
        }
            
        \foreach \i in {0,...,\numexpr\N-1\relax}{
          \draw[\latticeColor] (\i*\a+\LDx, -\margin+\LDy) -- (\i*\a+\LDx, \a*\N-\a + \margin+\LDy);
        }
        \node[blue!70] at (-0.3,0.25) {0};
        \node[blue!70] at (-0.3,0.75) {1};
        \node[blue!70] at (-0.3,1.25) {2};
        \node[blue!70] at (-0.3,1.75) {3};
        \node[blue!70] at (2.3,0.25) {3};
        \node[blue!70] at (2.3,0.75) {2};
        \node[blue!70] at (2.3,1.25) {1};
        \node[blue!70] at (2.3,1.75) {0};
        \node[red!70] at (0.25,-0.3) {0};
        \node[red!70] at (0.75,-0.3) {1};
        \node[red!70] at (1.25,-0.3) {2};
        \node[red!70] at (1.75,-0.3) {3};
        \node[red!70] at (0.25,2.3) {3};
        \node[red!70] at (0.75,2.3) {2};
        \node[red!70] at (1.25,2.3) {1};
        \node[red!70] at (1.75,2.3) {0};
    \end{scope}

    \end{tikzpicture}
    \caption{Identification patterns of the link variables for a torus, Klein bottle, and real projective plane are shown. The links across the boundaries with the same labels are connected.}
    \label{fig:manifolds_identification}
\end{figure}
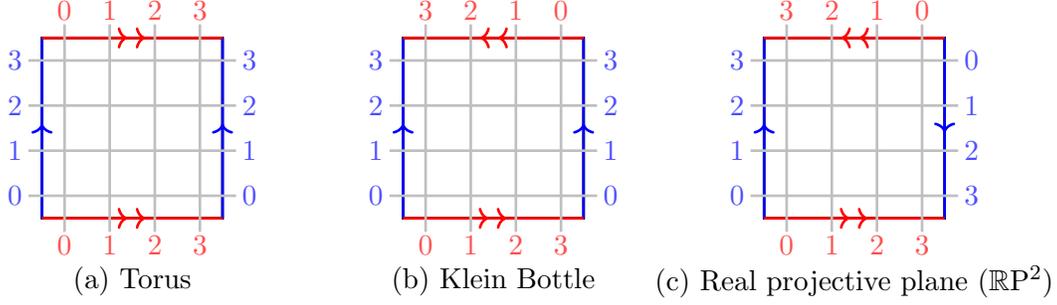

As a warm-up, let us consider the simplest example: a flat torus (see Fig.~\ref{fig:manifolds_identification} (a)).
For each of two directions ($x$ and $y$), there are two possible boundary conditions:
periodic ($P$) or antiperiodic ($A$).
Specifically they are given by
\begin{align}
\psi\left(N_xa,y\right) = \pm \psi\left(0,y\right), \label{eq:TorusBC_x}
 && \psi\left(x,N_y a\right) = \pm \psi\left(x,0\right), 
\end{align}
for arbitrary $x$ and $y$.
There are four different $\pinm$ structures on the torus, $PP$, $PA$, $AP$ and $AA$
where each alphabet denotes the boundary condition in the order of $x$ and $y$.

\subsection{Klein bottle $KB$}
The Klein bottle is obtained by gluing the upper and lower edges of 
a cylinder after reversing the orientation (see Fig.~\ref{fig:manifolds_identification} (b)). 
First we make a cylinder identifying $x=0$ and $x=N_x a$, 
with $P$ or $A$ boundary conditions:
\begin{align}
  \psi\left(N_x a,y\right) = \pm \psi\left(0,y\right) \label{eq:KBBC_x}.
\end{align}
Then we glue the $y = 0$ with $y = N_y a$ edges with 
a group operation of $\pinm(2)$ associated with the $O(2)$ operation that reverses the $x$-direction.
The operation is given by 
\begin{equation}
  \begin{alignedat}{4}
\psi\left(x,N_y a\right) &=& s R_x\psi(x,0)& =& s \gamma^1 \bar{\gamma}\psi\left((N_x-1)a-x,0\right),
\label{eq:KBBC_y}
  \end{alignedat}
\end{equation}
where we have two choices of the sign $s=\pm 1$. 
For the backward difference at the slice $y=0$, 
which has to be the conjugate of the forward difference, 
it is understood
that the inverse of the above boundary condition
\begin{equation}
  \begin{alignedat}{4}
\psi\left(x,-a\right) &=& s R_x^{-1}\psi(x,0),
  \end{alignedat}
\end{equation}
holds. Hence, the Klein bottle admits four distinct $\pinm$ structures we denote by 
$P+,P-,A+$ and $A-$, where the second $\pm$ corresponds to the sign $s$ of Eq.~\eqref{eq:KBBC_y}. 
A similar approach for considering fermion partition function on non-orientable manifolds is represented in \cite{Mages:2015scv}.

\subsection{Real projective plane $\RPt$}
\label{sec:RP2}
In order to consider Majorana fermions on an $\RPt$, we  glue both the top and bottom 
and the left and right boundaries of the lattice with a twist in each direction (see Fig.~\ref{fig:manifolds_identification} (c)).
Explicit forms of the boundary conditions are 
\begin{equation}
  \begin{alignedat}{4}
   \psi\left(N_x a,y\right) &=& s_x R_y\psi(0,y)& =& s_x \gamma^2 \bar{\gamma}\psi\left(0,(N_y-1) a-y\right)  \\
   \psi\left(x,N_y a\right) &=& s_y R_x\psi(x,0)& =& s_y \gamma^1 \bar{\gamma}\psi\left((N_x-1) a-x,0\right),
  \end{alignedat}\label{eq:rptBC}
\end{equation}
as well as their conjugates $\psi\left(-a,y\right) = s_x R_y^{-1}\psi(0,y)$ and 
$\psi\left(x,-a\right) = s_y R_x^{-1}\psi(x,0)$.
Here $s_y=\pm 1$ and $s_x=\pm 1$ can be taken independently.

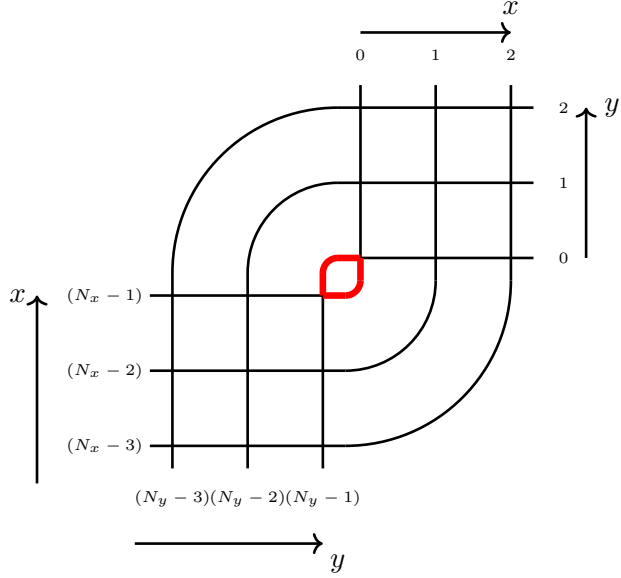
\begin{figure}[htbp]
    \centering
    \begin{tikzpicture}[scale=1, line width=1pt]

    \begin{scope}[xshift=0cm]
      \def\N{3}          
      \def\a{1}        
      \def\LDx{0}
      \def\LDy{0}
      \def\margin{0.3}  
      \def\latticeColor{black}
          
      \foreach \j in {0,...,\numexpr\N-1\relax}{
        \draw[\latticeColor] (-\margin+\LDx,\j*\a+\LDy)--(\N*\a-1*\a+\margin+\LDx,\j*\a+\LDy);
      }
          
      \foreach \i in {0,...,\numexpr\N-1\relax}{
        \draw[\latticeColor] (\i*\a+\LDx, -\margin+\LDy) -- (\i*\a+\LDx, \a*\N-\a + \margin+\LDy);
      }
      \node[font=\tiny] at (-0.9,\a*2){$(N_x-1)$};
      \node[font=\tiny] at (-0.9,\a*1){$(N_x-2)$};
      \node[font=\tiny] at (-0.9,\a*0){$(N_x-3)$};
      \node[font=\tiny] at (\a*2,-0.7){$(N_y-1)$};
      \node[font=\tiny] at (\a*1,-0.7){$(N_y-2)$};
      \node[font=\tiny] at (\a*0,-0.7){$(N_y-3)$};
      \draw[->] (-0.5,-1.3) -- (2,-1.3);
      \node[below] at (2.2,-1.3){$y$};
      \draw[->] (-1.8,-0.5) -- (-1.8,2);
      \node[left] at (-1.8,2){$x$};

      \def\LDx{2.5}
      \def\LDy{2.5}
          
      \foreach \j in {0,...,\numexpr\N-1\relax}{
        \draw[\latticeColor] (-\margin+\LDx,\j*\a+\LDy)--(\N*\a-1*\a+\margin+\LDx,\j*\a+\LDy);
      }
          
      \foreach \i in {0,...,\numexpr\N-1\relax}{
        \draw[\latticeColor] (\i*\a+\LDx, -\margin+\LDy) -- (\i*\a+\LDx, \a*\N-\a + \margin+\LDy);
      }

      \draw[->] (\LDx ,\LDy+2*\a+1) -- +(2,0);
      \node[above] at (\LDx +2,\LDy+2*\a+1+0.1){$x$};

      \draw[->] (\LDx +2*\a+1,\LDy) -- +(0,2);
      \node[right] at (\LDx +2*\a+1+0.1,\LDy+2){$y$};

      \node[font=\tiny] at (\LDx+\a*2+0.7,\LDy+\a*0){$0$};
      \node[font=\tiny] at (\LDx+\a*2+0.7,\LDy+\a*1){$1$};
      \node[font=\tiny] at (\LDx+\a*2+0.7,\LDy+\a*2){$2$};
      \node[font=\tiny] at (\LDx+\a*0,\LDy+\a*2+0.7){$0$};
      \node[font=\tiny] at (\LDx+\a*1,\LDy+\a*2+0.7){$1$};
      \node[font=\tiny] at (\LDx+\a*2,\LDy+\a*2+0.7){$2$};

      \draw[out=0,in=-90] (\a*2+\margin,0*\a) to (\LDx+2*\a,\LDy-\margin);
      \draw[out=0,in=-90] (\a*2+\margin,1*\a) to (\LDx+1*\a,\LDy-\margin);
      \draw[red,line width = 2.5pt,out=0,in=-90] (\a*2+\margin,2*\a) to (\LDx+0*\a,\LDy-\margin);
      \draw[red,line width = 2.5pt,out=0,in=-90] (\LDx+0*\a,\LDy-\margin) -- ++(0,\margin);
      \draw[red,line width = 2.5pt,out=0,in=-90] (\a*2+\margin,2*\a) -- ++(-\margin,0);
      \draw[out=90,in=180] (0*\a,\a*2+\margin) to (\LDx-\margin,2*\a+\LDy);
      \draw[out=90,in=180] (1*\a,\a*2+\margin) to (\LDx-\margin,1*\a+\LDy);
      \draw[red,line width = 2.5pt,out=90,in=180] (2*\a,\a*2+\margin) to (\LDx-\margin,0*\a+\LDy);
      \draw[red,line width = 2.5pt,out=0,in=-90] (2*\a,\a*2+\margin) -- ++(0,-\margin);
      \draw[red,line width = 2.5pt,out=0,in=-90] (\LDx-\margin,0*\a+\LDy) -- ++(\margin,0);
    \end{scope}

    \begin{scope}[xshift=8.0cm]
        
    \end{scope}

    \end{tikzpicture}
    \caption{The lattice network around the singularity of $\RPt$. Note that $(0,0)$ and 
$((N_x-1)a,(N_y-1)a)$ points are connected by two different link variables 
(denoted by thick ones), which form a nontrivial Wilson loop discussed in the main text.}
    \label{fig:RP2singularity}
\end{figure}

With the above boundary conditions, 
we have two pairs of singular points on the lattice.
One is $(x,y)=(0,0)$ and $((N_x-1)a,(N_y-1)a)$, 
which are connected by both of the $x$ and $y$ differences
since
\begin{align}
 a\nabla_x^f\psi\left((N_x-1) a,(N_y-1)a\right) &= s_x \gamma^2 \bar{\gamma}\psi\left(0,0\right)-\psi\left((N_x-1) a,(N_y-1)a\right),\nonumber\\
 a\nabla_y^f\psi\left((N_x-1)a,(N_y-1) a\right) &= s_y \gamma^1 \bar{\gamma}\psi\left(0,0\right)-\psi\left((N_x-1) a,(N_y-1)a\right).
 \label{eq:bccorner00}
\end{align}
Another pair is $(x,y)=((N_x-1)a,0)$ and $(0,(N_y-1)a)$ to which 
\begin{align}
  a\nabla_x^f\psi\left((N_x-1)a,0\right) &= s_x \gamma^2 \bar{\gamma}\psi\left(0,(N_y-1)a\right)-\psi\left((N_x-1)a,0\right),\nonumber\\
 a\nabla_y^f\psi\left(0,(N_y-1) a\right) &= s_y \gamma^1 \bar{\gamma}\psi\left((N_x-a)a,0\right)-\psi\left(0,(N_y-1) a\right),
 \label{eq:bccornerN0}
\end{align}
hold.
We understand this as a consequence of the fact that the $\RPt$ cannot be realized as globally flat manifold.
In fact, we can interpret that there are two different links from $((N_x-1)a,(N_y-1)a)$
in the $x$ direction and $y$ direction, which are curved to end up with 
the same point $(0,0)$ as shown in Fig.\ref{fig:RP2singularity}.
For the two forward difference operators at $((N_x-1)a,(N_y-1)a)$,
the two factors $U_x=s_x \gamma^2 \bar{\gamma}$ and $U_y=s_y \gamma^1 \bar{\gamma}$
in front of $\psi(0,0)$ in Eq.~\eqref{eq:bccorner00} 
can be understood as the corresponding link variables 
of the local Lorentz $SO(2)$ gauge field.
This interpretation similarly works for Eq.~\eqref{eq:bccornerN0}.

It is then interesting to consider the Wilson loop $W_1$ formed by 
the above nontrivial link variables starting from $(0,0)$ 
in the counter-clockwise direction, as well as $W_2$ from $(0,(N_ya-1)a)$.
We have
\begin{equation}
 W_1=W_2=U_x U_y^{-1}=-s_xs_y \gamma_1\gamma_2 = \exp\left[-\pi s_x s_y T\right],
\end{equation}
with the local Lorentz generator $T=[\gamma_1,\gamma_2]/4$.
If our lattice setup is a good regularization of 
a smooth continuum $\RPt$, 
the Riemann curvature may be identified from $W_{1,2}$ as 
$R_1=R_2=-\pi s_x s_y/A$, where $A$ is the area of each Wilson loop.
Then we can compute the Euler characteristic of the lattice by
\begin{equation}
 \chi = \frac{1}{2\pi}\int R = \frac{(R_1+R_2)A}{2\pi} = -s_x s_y.
\end{equation}
Since $\chi$ of $\RPt$ is unity, our assumption of the smoothness requires $s_x s_y=-1$.
Therefore, we expect that two choices $(s_x,s_y)=(+,-)$ and $(-,+)$ will represent
the two $\pinm$ structure of $\RPt$, while the other $(+,+)$ and $(-,-)$ seem to fail
in regulating the singularity.
 
\subsection{Introducing cross-caps}
\label{sec:cross-cap}

\begin{figure}[htbp]
\centering
\begin{tikzpicture}

\coordinate (A1) at (0,0);
\coordinate (A2) at (3,0);
\coordinate (A3) at (3,3);
\coordinate (A4) at (0,3);

\foreach \i/\j in {1/2, 2/3, 3/4, 4/1} {
  \draw[
    thick,
    decorate,
    decoration={snake, amplitude=3pt, segment length=60pt}
  ] (A\i) -- (A\j);
}
\node[font=\scriptsize, yshift=6pt] at (0.5,1.5) {$(x_L,y_c)$};
\node[font=\scriptsize, yshift=6pt] at (2.5,1.5) {$(x_R,y_c)$};
\node[font=\small] at (1.5,-0.6) {(a) Cutting the manifold};
\draw (0.5,1.5)--(2.5,1.5);

\begin{scope}[xshift=4.5cm]
  \coordinate (B1) at (0,0);
  \coordinate (B2) at (3,0);
  \coordinate (B4) at (0,3);
  \coordinate (B3) at (3,3);
  \foreach \i/\j in {1/2, 2/3, 3/4, 4/1} {
    \draw[
      thick,
      decorate,
      decoration={snake, amplitude=3pt, segment length=60pt}
    ] (B\i) -- (B\j);
  }
  \draw[thick,->-] (0.5,1.5) to [out=45, in=135] (2.5,1.5);
  \draw[thick,->-] (2.5,1.5) to [out=-135,in=-45] (0.5,1.5);
  \node[font=\small, yshift=9pt, align=center] at (1.5,-1.25) {(b) and identifying edges \\ reversing the orientation};
\end{scope}

\begin{scope}[xshift=9.0cm]
  \coordinate (C2) at (3,0);
  \coordinate (C1) at (0,0);
  \coordinate (C4) at (0,3);
  \coordinate (C3) at (3,3);
  \foreach \i/\j in {1/2, 2/3, 3/4, 4/1} {
    \draw[
      thick,
      decorate,
      decoration={snake, amplitude=3pt, segment length=60pt}
    ] (C\i) -- (C\j);
  }
  \coordinate (L) at (0.5,1.5);  
  \coordinate (R) at (2.5,1.5);  
  \coordinate (M) at (1.5,1.5);  

  \path[name path=upperEye] (L) to[out=45,in=135] (R);
  \path[name path=lowerEye] (R) to[out=-135,in=-45] (L);
  \draw[thick] (L) to [out=45, in=135] (R);
  \draw[thick] (R) to [out=-135,in=-45] (L);
  \foreach \ang in {30,89.9,150} {
    \path[name path=rayUp] (M) -- ++(\ang:2);
    \path[name intersections={of=rayUp and upperEye, by=iup}];

    \draw[thin] (M) -- (iup);
  }

  \foreach \ang in {-30,-90.1,-150} {
    \path[name path=rayLow] (M) -- ++(\ang:2);
    \path[name intersections={of=rayLow and lowerEye, by=ilow}];
    \draw[thin] (M) -- (ilow);
  }
  \node[font=\small] at (1.5,-0.6) {(c) forms a cross-cap.};
\end{scope}

\end{tikzpicture}
\caption{Introducing a crosscap.}
\label{fig:crosscap-intro}
\end{figure}
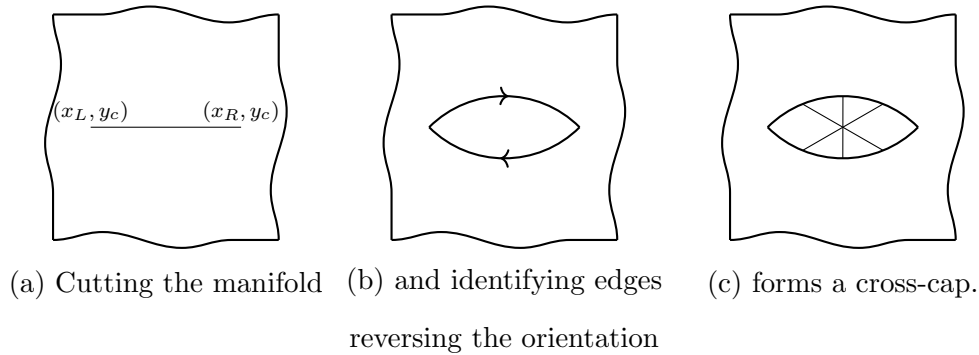

Introducing cross-caps, we can consider more complicated manifolds.
For instance, adding cross-cap to a torus yields a manifold whose topology is a triple connected sum of $\RPt$ which we denote by ${\#}^3\RPt$.
A cross cap is obtained by cutting the manifold along a line segment and identifying the two edges with reversed orientation
as Fig.~\ref{fig:crosscap-intro}.

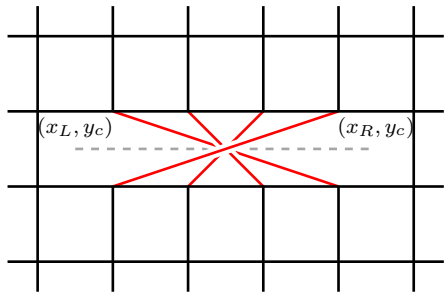
\begin{figure}[htbp]
    \centering
    \begin{tikzpicture}[scale=1, line width=1pt]

      \def\Nx{4}          
      \def\Ny{6}
      \def\a{1}        
      \def\LDx{0}
      \def\LDy{0}
      \def\margin{0.4}  
      \def\latticeColor{black}
      \def\latticeColorB{red}

      \draw[dashed,gray!70] (0.5,1.5) -- (4.5,1.5);
      \node[above,font=\scriptsize] at (0.5,1.5) {$(x_L,y_c)$};
      \node[above,font=\scriptsize] at (4.5,1.5) {$(x_R,y_c)$};

      \draw[\latticeColor] (0*\a+\LDx, \a*1 +\LDy) --(0*\a+\LDx, 2+\LDy) ;
      \draw[\latticeColor] (5*\a+\LDx, \a*1 +\LDy) --(5*\a+\LDx, 2+\LDy) ;
      \draw[\latticeColorB,preaction={draw, line width=4pt, white}] (4*\a+\LDx, \a*1 +\LDy) --(1*\a+\LDx, 2+\LDy) ;
      \draw[\latticeColorB,preaction={draw, line width=4pt, white}] (3*\a+\LDx, \a*1 +\LDy) --(2*\a+\LDx, 2+\LDy) ;
      \draw[\latticeColorB,preaction={draw, line width=4pt, white}] (2*\a+\LDx, \a*1 +\LDy) --(3*\a+\LDx, 2+\LDy) ;
      \draw[\latticeColorB,preaction={draw, line width=4pt, white}] (1*\a+\LDx, \a*1 +\LDy) --(4*\a+\LDx, 2+\LDy) ;
      
      \foreach \j in {0,...,\numexpr\Nx-1\relax}{
        \draw[\latticeColor] (-\margin+\LDx,\j*\a+\LDy)--(\Ny*\a-1*\a+\margin+\LDx,\j*\a+\LDy);
      }
          
      \foreach \i in {0,...,\numexpr\Ny-1\relax}{
        \draw[\latticeColor] (\i*\a+\LDx, -\margin+\LDy) -- (\i*\a+\LDx, \a*2-\a +\LDy);
        
        \draw[\latticeColor] (\i*\a+\LDx, 2+\LDy) -- (\i*\a+\LDx, \a*\Nx-\a + \margin+\LDy);
      }

    \end{tikzpicture}
    \caption{Lattice network around the crosscap.}
    \label{fig:crosscapLattice}
\end{figure}

Let us consider a line segment $(x_L,y_c)$ and $(x_R,y_c)$.
In the continuum case, a cross-cap is obtained by gluing two patches ,the lower ($y\leq y_c$) and the upper ($y\geq y_c$), by the following boundary conditions 
\begin{equation}
\label{eq:crosscap}
  \psi_{\mathrm{lower}} (x,y_c) = \begin{cases}
  \pm \gamma^1 \bar{\gamma}\psi_{\mathrm{upper}} (x_L+x_R-x,y_c) & \text{if } x_L < x < x_R \\
  \psi_{\mathrm{upper}} (x,y_c) & \text{otherwise}
  \end{cases}.
\end{equation}
Both choices of the sign $\pm$ is allowed, and hence the introduction of 
one cross-cap doubles the number of possible $\pinm$ structures.
To realize this construction on the lattice, we impose the rule that, when the forward difference operator $\nabla_y^f$ refers to a point across the crosscap segment from the point $(x,y)$ ({\it i.e.}, the crosscap lies between $(x,y)$ and $(x,y+1)$), it refers to $\pm \gamma^1 \bar{\gamma}\psi (x_L+x_R-x,y+1)$ instead of $\psi (x,y+1)$ and vice versa for the opposite direction. This lattice version of the cross-cap is presented in Fig.~\ref{fig:crosscapLattice}.

\subsection{Including domain-walls}
\label{sec:DWAPS}

In order to consider the ABK (or $\eta$ invariant in more general setups) on open manifolds,
let us introduce a  domain-wall fermion mass term \cite{Kaplan:1992bt,Shamir:1993zy}.
We consider a closed continuum subspace $X$ on the periodic or anti-periodic lattice, 
and  the  position-dependent mass is defined by
\begin{equation}
  m_X(x,y) = \begin{cases}
    -m_0 & \text{if } (x,y) \text{ in } X, \\
    +m_0 & \text{otherwise}
  \end{cases},
\end{equation}
where we assume that $m_0>0$. 
In the conventional lattice simulations, one often takes $X$ with flat boundaries
but here we allow the cases with curved domain-walls \cite{Aoki:2022cwg, Aoki:2022aez,Kaplan:2023pvd,Kaplan:2023pxd,Aoki:2024bwx,Clancy:2024bjb}.

We define the lattice version of the ABK invariant on open manifolds by 
\begin{equation}
 \beta^{\mathrm{lat}}(m_X,a;X):= \frac{4}{\pi i}\arg \left( \frac{\mathrm{Pf}\left( C D_{W}(m_X) \right)}{\mathrm{Pf}\left( C D_{W}(m_0)\right)} \right).
\end{equation}
If our lattice calculation of the ABK invariant is consistent with the continuum result,
the values depend only on the topology of the subregion $X$ and not on the local properties in the $a\to 0$ and $m_0 \to \infty$ limits as in the closed manifold case.
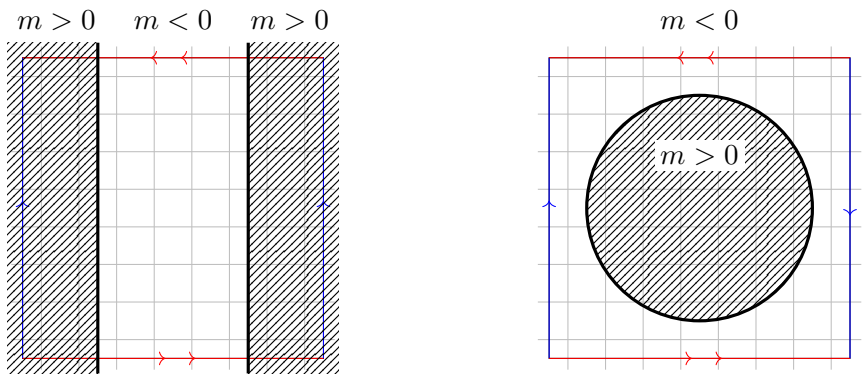
\begin{figure}[htbp]
    \centering
    \begin{tikzpicture}
   
        \def\N{8}          
        \def\a{0.5}        
        \def\LDx{0.25}
        \def\LDy{0.25}
        \def\margin{0.4}  
        \def\latticeColor{gray!50}
            
        \foreach \j in {0,...,\numexpr\N-1\relax}{
          \draw[\latticeColor] (-\margin+\LDx,\j*\a+\LDy)--(\N*\a-1*\a+\margin+\LDx,\j*\a+\LDy);
        }
            
        \foreach \i in {0,...,\numexpr\N-1\relax}{
          \draw[\latticeColor] (\i*\a+\LDx, -\margin+\LDy) -- (\i*\a+\LDx, \a*\N-\a + \margin+\LDy);
        }

        \draw (0,0) rectangle (4,4);

        \draw[->-,blue] (0,0) -- (0,4);
        \draw[->-,blue] (4,0) -- (4,4);

        \draw[->>-,red] (0,0) -- (4,0);
        \draw[->>-,red] (4,4) -- (0,4); 
        \fill[pattern=north east lines, pattern color=black] (-0.2,-0.2) rectangle (1,4.2);
        \fill[pattern=north east lines, pattern color=black] (3,-0.2) rectangle (4.2,4.2);
        \draw[very thick] (1,-0.2) -- (1,4.2);
        \draw[very thick] (3,-0.2) -- (3,4.2);
        \node[fill=white, inner sep=2pt,font=\small] at (0.45,4.5) {$m>0$};
        \node[fill=white, inner sep=2pt,font=\small] at (2,4.5) {$m<0$};
        \node[fill=white, inner sep=2pt,font=\small] at (3.55,4.5) {$m>0$};
        \begin{scope}[xshift=7cm]
          \def\N{8}          
          \def\a{0.5}        
          \def\LDx{0.25}
          \def\LDy{0.25}
          \def\margin{0.4}  
          \def\latticeColor{gray!50}
              
          \foreach \j in {0,...,\numexpr\N-1\relax}{
            \draw[\latticeColor] (-\margin+\LDx,\j*\a+\LDy)--(\N*\a-1*\a+\margin+\LDx,\j*\a+\LDy);
          }
              
          \foreach \i in {0,...,\numexpr\N-1\relax}{
            \draw[\latticeColor] (\i*\a+\LDx, -\margin+\LDy) -- (\i*\a+\LDx, \a*\N-\a + \margin+\LDy);
          }
  
          \draw (0,0) rectangle (4,4);
  
          \draw[->-,blue] (0,0) -- (0,4);
          \draw[->-,blue] (4,4) -- (4,0);
  
          \draw[->>-,red] (0,0) -- (4,0);
          \draw[->>-,red] (4,4) -- (0,4); 
          \filldraw[pattern=north east lines, pattern color=black,very thick] (2,2) circle [radius=1.5];

          \node[fill=white, inner sep=2pt,font=\small] at (2,4.5) {$m<0$};
          \node[fill=white, inner sep=2pt,font=\small] at (2,2.7) {$m>0$};
        \end{scope}
    \end{tikzpicture}
    \caption{ In order to realize a M\"obius strip, we consider
the two types of the domain-wall mass term shown above.
The unshaded regions have a negative mass, while the
shaded regions are assigned to have a positive mass. The unshaded regions possess the M\"obius strip topology.
The left panel represents two domain-walls to form a straight M\"obius strip on a Klein bottle,
and the right one corresponds to a circular domain-wall removing a disk region from an $\RPt$, 
which forms a curved M\"obius strip.}
    \label{fig:mobiusDW}
\end{figure}

A typical example of two-dimensional non-orientable open manifolds is the Möbius strip, 
which can be identified as a half of the Klein bottle discussed above.
The corresponding domain-wall mass term is given by 
\begin{equation}
  m_{\mathrm{MS}}(x,y) = \begin{cases}
    -m_0 & \text{if } \frac{(N_y+1) a -w}{2}<y< \frac{(N_y+1) a +w}{2}, \\
    +m_0 & \text{otherwise}
  \end{cases}, \label{eq:DWmassMS}
\end{equation}
where $w$ is the width and we set the same boundary condition as Eq.~\eqref{eq:KBBC_y}.
Another way to realize a Möbius strip is to remove a disc from an $\RPt$. When we introduce a domain-wall mass term 
\begin{equation}
  m_{\overline{\mathrm{disk}}}(x,y) = \begin{cases}
    +m_0 & \text{if } \sqrt{(x-x_0)^2+(y-y_0)^2} \leq r, \\
    -m_0 & \text{otherwise}
  \end{cases}. \label{eq:DWmassMS2}
\end{equation}
The two domain-wall configurations used to realize the Möbius strip are illustrated in Fig.~\ref{fig:mobiusDW}.
The subregion outside of the disk ($\overline{\mathrm{disk}}$) has the same topology as the Möbius strip introduced in Eq.~\eqref{eq:DWmassMS}. 
If our formulation of the AKB invariant is topological,
these two different lattice geometry should give the same value of $\beta^{\mathrm{lat}}(m_X,a;X)$ in the $a\to 0$ and $m_0 \to \infty$ limits.

Now we are ready to compute $\Pf(C D_W(m_X))$ with or without domain-walls
to extract the complex phase on various types of non-orientable manifolds discussed above.
Some of them have translational symmetry so that we can analytically evaluate the Pfaffian 
via Fourier transformation, which will  be discussed in the next section.
For more general manifolds, we will perform numerical evaluation on 
a finite lattice with $N_x, N_y\lesssim 30$ in Sec.~\ref{ss:numerical_calculation}.

\section{Fourier analysis on flat closed manifolds}
\label{sec:Fourier}
Since the Wilson-Dirac operator $D_W(m)$ is a finite complex matrix,
it is not difficult to compute, at least numerically,
the partition function $Z=\Pf(CD_W(m))$. 
For the torus and Klein bottle cases defined in the previous section,
we can analytically perform the program thanks to the translational symmetry.
By the Fourier transformation, we can microscopically identify
specific eigenmodes that contribute to the phase of $Z$ and thus to the ABK invariant.

In the momentum space, the Wilson-Dirac operator $D_W(m)$ is expressed as
\begin{equation}
   i\gamma^1 \frac{1}{a}\sin(a p_x)+i\gamma^2 \frac{1}{a}\sin(a p_y)+ m + \frac{1}{a} \left( 2 -\cos(a p_x)-\cos(ap_y) \right) ,
\end{equation}
and it has eigenvalues
\begin{equation}
  \label{eq:eigenvalues}
  \mu^{\pm}(p_x,p_y)=\pm i \frac{1}{a}\sqrt{\sin(a p_x)^2+\sin(a p_y)^2} + m + \frac{1}{a} \left( 2-\cos(a p_x)-\cos(a p_y) \right),
\end{equation}
with eigenvectors
\begin{equation}
  u^\pm (x,y;p_x,p_y) =  \frac{1}{\sqrt{2}} \begin{pmatrix}1 \\  \pm  \frac{\sin(a p_x)+i\sin(a p_y)}{\sqrt{\sin(a p_x)^2+\sin(a p_y)^2}}\end{pmatrix} e^{i\left( p_x x + p_y y \right)}. \label{eq:latt_eigvec_torus}
\end{equation}
The boundary conditions on the target manifold constrain
the allowed values of the momentum and form of the eigenfunctions.

\subsection{Flat torus $T^2$}
In continuum theory, it is known for the two-dimensional torus $T^2$
that among the four $\pinm$ structures, only the $PP$ boundary condition gives
a nontrivial phase $(-1)$ to the partition function;
in other words, the ABK invariant $\beta(PP)=4$,
while the others ($PA,AP,AA$) have $\beta=0$.
This $\ZZ_2$ structure reflects the fact that $T^2$ is a orientable spin manifold and
the ABK invariant reduces to the standard mod-two index. These results can be checked straightforwardly as follows, in contrast to the nontriviality of the Klein-bottle case discussed in the next section.

Let us examine if our lattice set up reproduces this simplest result.
Under the boundary conditions \eqref{eq:TorusBC_x}, 
The allowed momenta are
\begin{equation}
  p_{x/y} = \begin{cases}
  \frac{2\pi}{aN_{x/y}}k & (k=0,\dots,N_{x/y}) \quad \text{  if periodic,} \\
  \frac{2\pi}{aN_{x/y}}\left(k+\frac{1}{2}\right) & (k=0,\dots,N_{x/y}) \quad \text{  if antiperiodic}
  \end{cases},
\end{equation}
and the $u^\pm (x,y;p_x,p_y)$ set forms a linearly independent basis.
The partition function can be written as
\begin{equation}
  Z = \Pf(CD_W(m))={\prod_{p_x,p_y,\pm}^{}}'\mu^{\pm}(p_x,p_y),
\end{equation}
where ${\prod}'$ takes one eigenvalue from the doubly degenerate eigenvalues.

When $\sin(a p_x),\sin(a p_y) \ne 0$, i.e., $p_x,p_y \ne 0 \ \text{and} \ne \pi/a $,
the eigenvalues appear in complex conjugate pairs.
Therefore these eigenvalues do not contribute to the phase of the partition function.
In addition, when $p_x$ or $p_y$ is $\pi/a $ (corresponding to the doubler modes) and $-2<ma$, the eigenvalues become positive and do not contribute to the phase. The only nontrivial  contribution to the phase comes from the (one of doubly degenerate) $p_x=p_y=0$ eigenvalue $\mu(0,0)=m$ and this eigenvalue exist only when the $\pinm$ structure is $PP$.

In the $PP$ case, the ratio of the partition function in the topological phase
($-2<ma<0$) to that in the trivial vacuum ($0<M$) exhibits the phase
\begin{align}
  \frac{Z(m;PP)}{Z(M;PP)} \propto \frac{m}{M} \propto -1, \\
  \beta^{\mathrm{latt}}(PP)=4.
\end{align}
This result is consistent with the continuum theory.
The feature that the phase arises solely from the zero mode of
massless Dirac operator (mod-two index) is also reproduced.

\subsection{Klein bottle $KB$}
\label{sec:KBFourier}
On a Klein bottle, the nontrivial boundary condition given in
Eqs.~\eqref{eq:KBBC_x} and \eqref{eq:KBBC_y}
mixes the modes with $\pm p_x$ to form the eigenvectors  
\begin{equation}
    u^\pm_{\mathrm{KB}} (x,y;p_x,p_y)=\frac{1}{2} u^\pm (x,y;p_x,p_y) + s \frac{1}{2} i (-1)^{k_y}  \frac{\sin(a p_x)+i\sin(a p_y)}{\sqrt{\sin(a p_x)^2+\sin(a p_y)^2}}  u^\pm (x,y;-p_x,p_y), \label{eq:eigenFunction_KB}
\end{equation}
with the eigenvalues $\mu^\pm (p_x,p_y)$ in Eq.~\eqref{eq:eigenvalues}
, where $s=\pm$ is the sign which corresponds to the choice of the boundary conditions $Ps,As$,
and the allowed momenta are given by
\begin{align}
  p_{x} &= \begin{cases}
  \frac{2\pi}{N_{x} a}k_x & (k_x=0,\dots,\lfloor \frac{N_x}{2} \rfloor) \quad \text{  if periodic} \\
  \frac{2\pi}{N_{x} a}\left(k_x+\frac{1}{2}\right) & (k_x=0,\dots,\lfloor \frac{N_x-1}{2} \rfloor) \quad \text{  if antiperiodic },
  \end{cases} \label{eq:KBpxs}\\
  p_y &= \frac{\pi}{N_y a}\left(k_y+\frac{1}{2}\right) \quad (k_y=0,\dots,2N_y-1).  \label{eq:KBpys}
\end{align}
To derive this result, it is useful to start from the eigenfunctions on a torus with twice the length $2N_ya$ in the $y$-direction which is the double cover of the Klein bottle. We impose the anti-periodic boundary condition in the $y$ direction. The Klein bottle is identified as a $\ZZ_2$ quotient of this torus.
The above eigenfunctions are obtained by projecting those on the torus onto the $\ZZ_2$-invariant subspace.
See appendix \ref{app:KB_eigenvectores} for the details.

For $p_x \ne 0$ and $\ne\pi/a$ , the eigenfunctions $u^+_{\mathrm{KB}}$ and $u^-_{\mathrm{KB}}$ are independent, and therefore
their eigenvalues $\mu^+(p_x,p_y)$ and $\mu^-(p_x,p_y)= \mu^+(p_x,p_y)^*$
do not contribute to the complex phase of the partition function.
At $p_x=0$ or $=\pi/a$, the situation is different.
Substituting these values into Eq.~\eqref{eq:eigenFunction_KB}, one finds that either $u^{+}_{\mathrm{KB}}$ or $u^{-}_{\mathrm{KB}}$ is absent.
The $\pm$ label of surviving eigenfunctions $u^\pm_{\mathrm{KB}}$ alternate with $k_y$. 
More precisely, in the $P+$ and $A+$ case, surviving eigenfunctions are $u^-_{\mathrm{KB}},u^+_{\mathrm{KB}},u^-_{\mathrm{KB}},\dots$, for $k_y=0,1,2,\dots$, respectively, whereas for $P-$ and $A-$ case, the opposite sequence appears.

Therefore, if $p_x=0$ or $p_x=\pi/a$ appear in Eq.~\eqref{eq:KBpxs}, the corresponding factors
\begin{align}
  {\prod_{p_y}^{}}'\mu^{s(-1)^{k_y+1}}(0,p_y), &&  {\prod_{p_y}^{}}'\mu^{s(-1)^{k_y+1}}(\pi/a,p_y), \label{eq:KBfactors}
\end{align}
contribute to the phase of the partition function.
To evaluate these factors, we set
\begin{equation}
  \theta \left( p_y ;ma\right)  \equiv \arg(\mu^+(0,p_y)) = \arg\left(  + i \sin(a p_y) + ma + 1 -\cos(a p_y) \right).
\end{equation}
Using this function $\theta$, the complex phase of Eq.~\eqref{eq:KBfactors} can be rewritten as,
\begin{align}
  \exp \left( -s \frac{i}{2}\sum_{p_y} (-1)^{k_y}\theta \left( p_y ;ma\right) \right), &&
  \exp \left( -s \frac{i}{2}\sum_{p_y} (-1)^{k_y}\theta \left( p_y ;ma+2\right) \right) .\label{eq:KBfactors_2}
\end{align}
Here, the factor $1/2$ arises from taking only half of the degenerate eigenvalues. Note that there is no eigenvalues at $p_y=0,2\pi/a$ and no real eigenvalue, so that $\theta$ can be taken in range $-\pi < \theta(p_y;m) < \pi$ and there is no ambiguity in multiplying $\pm 1/2$ to it.

It is not difficult to show that the contribution from
$p_x=\pi/a$ can be neglected in the continuum limit.
Therefore, we can express the ABK invariant for the $P\pm$ case by a finite sum:
\begin{equation}
  \label{eq:betaKBFourier}
  \begin{aligned}
    \beta^{\rm latt}(Ps; m, a) &= -\frac{2s}{\pi} \sum_{p_y} (-1)^{k_y}\theta \left( p_y ;ma\right)
    \\ &= -\frac{2s}{\pi}\sum_{n=0}^{N_y-1} \left\{ \theta\left(  \frac{\pi}{a N_y }\left(2n+\frac{1}{2}\right);ma \right) -\theta\left(  \frac{\pi}{a N_y }\left(2n+1+\frac{1}{2}\right) ;ma\right) \right\}.
  \end{aligned}
\end{equation}
In the continuum limit $N_y\to \infty$, together with the $m\to \infty$ limit fixing $ma$ finite,
the summation can be evaluated as 
\begin{equation}
  \begin{aligned}
    \sum_{p_y} (-1)^{k_y}\theta \left( p_y ;ma\right) 
    &= \sum_{n=0}^{N_y-1} \left\{ -\theta' \left(  \frac{\pi}{a N_y }\left(2n+1\right);ma \right) \frac{\pi}{a N_y } +O\left(\frac{1}{N_y^2}\right) \right\}  \\
    &\xrightarrow[N_y\to\infty]{} - \frac{1}{2}\int_{0}^{\frac{2\pi}{a}} \theta' \left(  p_y;ma \right) dp_y\\
    &= \frac{\theta(0;ma)-\theta(2\pi/a;ma)}{2}. \\
  \end{aligned}
\end{equation}
When $m>0$, $\theta(0;ma)-\theta(2\pi/a;ma)$ makes no phase contribution because $\theta(0;ma)=\theta(2\pi/a;ma)=0$. On the other hand, for $m<0$, the $e^{i\theta(p_y;ma)}$ moves clockwise on the complex plane around the origin from $e^{i\theta(0;ma)}=-1$ to $e^{i\theta(2\pi/a;ma)}=-1$, and hence the phase difference is 
$\theta(0;ma)-\theta(2\pi/a;ma) = 2\pi$.
Thus, we can neglect the $p_x=\pi/a$ case, as well as positively massive fermion
phase.

Summarizing the above result, in the limit $a\to 0, m \to \infty$ and $N_{x/y} \to \infty$ , with $N_{x/y} a$ and $ma$ fixed, we have
\begin{align}
  Z(m;P\pm) &\propto \begin{cases}
    1 & \text{if } (ma>0),\\
    \exp \left(  \mp i\frac{\pi}{2} \right) = \mp i & \text{if } (-2<ma<0)
  \end{cases}&,\\
  Z(m;AP\pm) &\propto 1 \quad \text{if } (-2<ma),
\end{align}
and the corresponding lattice ABK invariant $\beta^{\mathrm{latt}}=\frac{4}{\pi i} \arg \frac{Z(m<0)}{Z(|m|)}$, becomes
\begin{align}
  \beta^{\mathrm{latt}}(m\to \infty,a\to 0;P\pm) &= \mp 2, \\ 
  \beta^{\mathrm{latt}}(m\to \infty,a\to 0;A\pm) &= 0.
\end{align}
This result is consistent with the continuum theory.

\begin{table}[htbp]
\centering
\caption{Numerical evaluation of the lattice ABK invariant on the Klein bottle $\beta^{\mathrm{latt}}(P+)$ at 
$|ma|=1$ and its error from the continuum value $\beta=-2$.}
\label{tab:KB}
\begin{tabular}{@{}c cc@{}} 
\toprule
$N_x\times N_y$ & $\beta^{\mathrm{latt}}(m,a;P+)$ & error $|\beta^{\mathrm{latt}}(P+)-\beta(P+)|$  \\ 
\midrule
$10 \times 10$ & $-1.99751\dots$ & $2.49 \times 10^{-3}$  \\ 
$20 \times 20$ & $-1.99999757\dots$ & $2.43 \times 10^{-6}$  \\ 
$30 \times 30$ & $-1.99999999762\dots$ & $2.37 \times 10^{-9}$ \\ 
\bottomrule
\end{tabular}
\end{table}

In Tab.~\ref{tab:KB}, we give numerical values of $\beta^{\mathrm{latt}}(m,a;P+)$
with various lattice spacings.
We can see that the value quickly converges to the
known continuum results, with errors which is exponentially
suppressed as $N_{x/y}$ increase.

\section{Numerical analysis on general manifolds} \label{ss:numerical_calculation}
In two-dimensional lattice gauge theory, it is numerically feasible 
to directly
evaluate the Pfaffian\footnote{We use the Julia package SkewLinearAlgebra.jl for our numerical computation in this section.} 
of the Wilson Dirac operator with a reasonable lattice size $N<30$.
Setting $m=-1/a$ on isotropic lattices with $N_x=N_y=N$ in the range $4\le N\le 24$,
we compute the ratio of the Wilson Dirac Pfaffian
and that of the Pauli-Villars contribution, from which
the  ABK invariant $\beta^\mathrm{latt}$ is extracted.
Fixing the physical volume size $L=Na$, increasing $N$
corresponds to approaching the continuum limit $a \to 0$
together with the large mass limit $m=1/a \to \infty$, at the same time.

Using the setup for the boundary condition described in Sec.~\ref{ss:lattice_setup},
we compute the ABK invariant on various manifolds with various $\pinm$ structures.

\begin{figure}[htbp]
  \centering
  \includegraphics[width=0.8\linewidth]{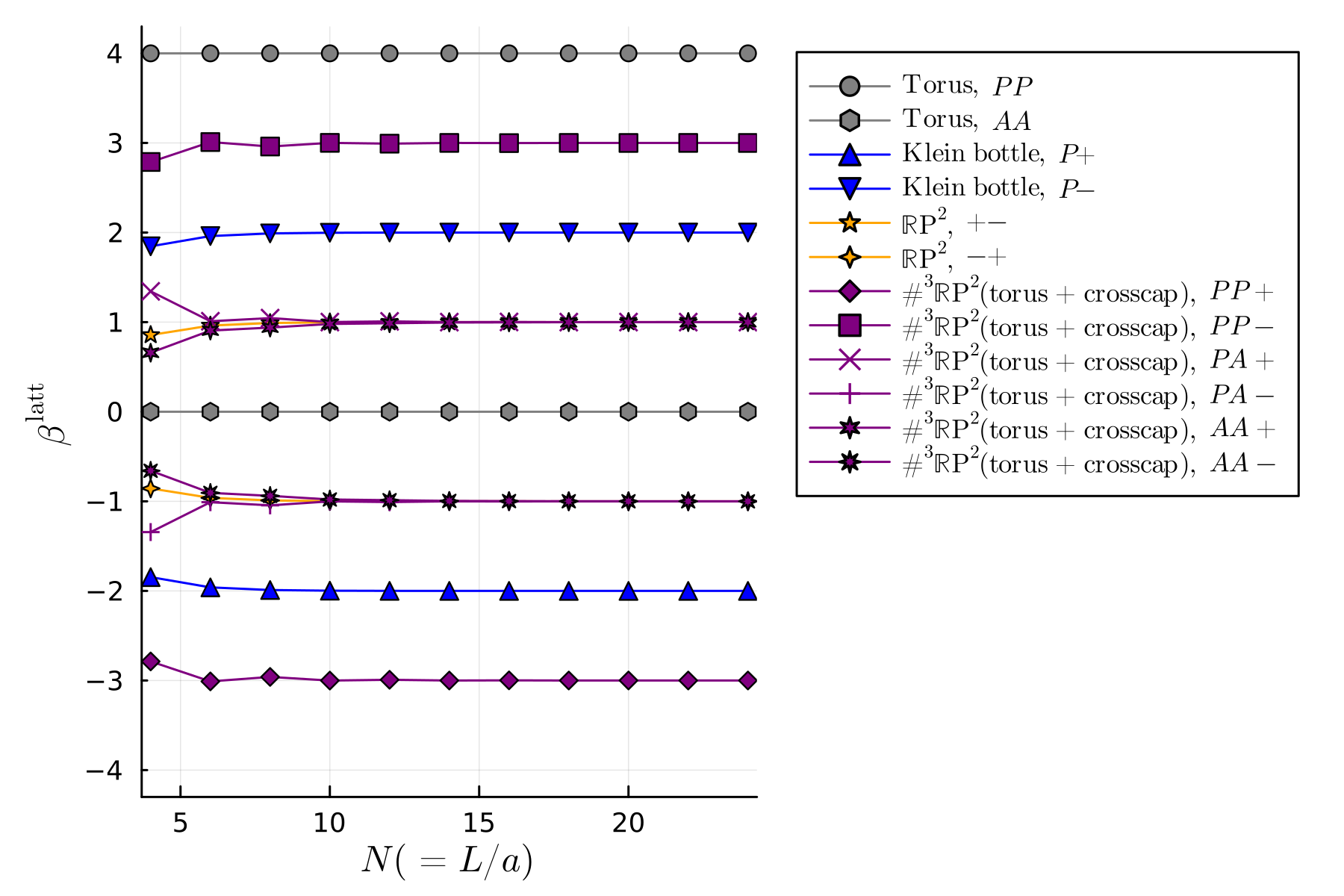} 
  \caption{The lattice ABK invariant $\beta^\mathrm{latt}$ is plotted as 
a function of $N=1/a$ for various closed manifolds.}
  \label{fig:batas_closed}
\end{figure}

First we consider the following closed manifolds:
a torus ($T^2; PP,AP,PA,AA$), Klein bottle ($KB;P\pm, A\pm$), real projective plane ($\RPt; \pm\mp$),
triple connected sum of $\RPt$ or ${\#}^3\RPt$.
For ${\#}^3\RPt$, it is constructed by introducing a crosscap
given by a twisted gluing explained in Sec.~\ref{sec:cross-cap} along a horizontal line segment
of the length $L/2$ at the center of the torus. 
Together with four choices of the boundary conditions in the $x$ and $y$ directions 
and the sign choice in Eq.~\ref{eq:crosscap},
there exist eight different $\pinm$ structures. 
Figure~\ref{fig:batas_closed} represents our numerical results for the various manifolds above.
It is obvious from the plot that the values of $\beta^\mathrm{latt}$ are quantized
into eight discrete values, which becomes stable already at $N\sim 10$.
All the obtained results at large $N$ are consistent with the known values in the continuum theory.
It is remarkable that the data for $\RPt$ and  ${\#}^3\RPt$ show correct values
even though the curvature on them is almost singular, or localized in some small-scale regions of the size $a$. 

For $\RPt$, we also try the ``incorrect'' boundary conditions $({+}{+})$ and $({-}{-})$ in Eq.~\eqref{eq:rptBC}.
We find that they give very small values $|\beta^{\mathrm{latt}}(\RPt,++)|,|\beta^{\mathrm{latt}}(\RPt,--)|<10^{-9}$,
which do not match any values of the ABK invariant on $\RPt$ in the continuum theory.
As discussed in Sec.~\ref{sec:RP2} these cases would contain singularities which
cannot be smoothed in the continuum limit, while the correct $({+}{-})$ and $({-}{+})$
cases seem to be successful in reproducing strong but finite curvatures on the two plaquettes
at the corners.

\begin{figure}[htbp]
  \centering
  \includegraphics[width=0.8\linewidth]{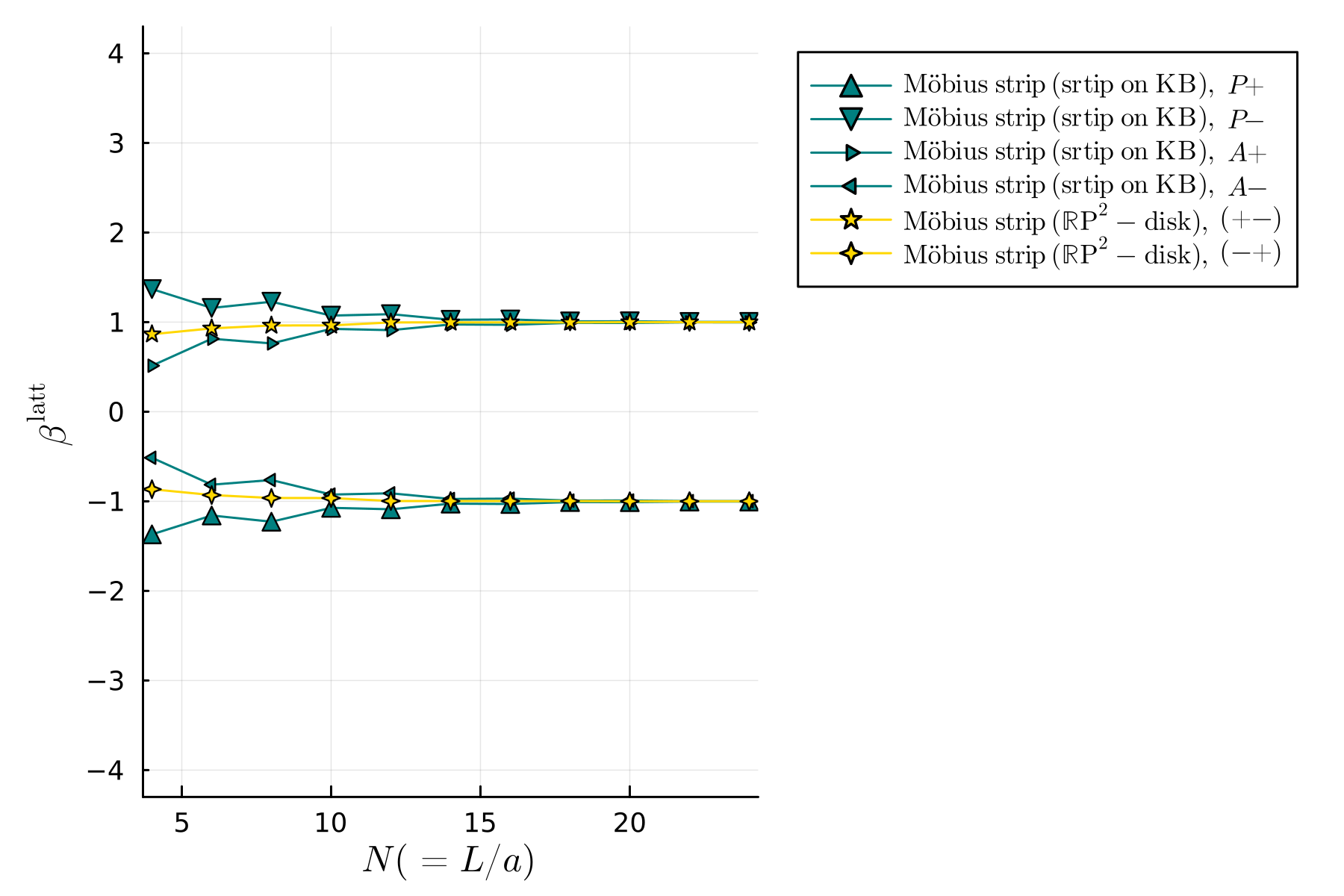} 
  \caption{The lattice ABK invariant $\beta^\mathrm{latt}$ is plotted 
as a function of $N=1/a$ for various setups of a Möbius strip.}
  \label{fig:batas_MS}
\end{figure}

Next we introduce the domain-wall mass term.
If our conjecture in Sec.~\ref{sec:DWAPS} is correct, 
the Pfaffian of the domain-wall fermion Dirac operator 
gives the ABK invariant of the negative mass region with boundaries 
located at the domain-wall. 
Here we examine this on M\"obius strips, 
for which the continuum value is $\beta=\pm 1$,
with two different setups of the domain-walls.
One is to put two straight domain-walls on a Klein bottle (given by Eq.~\eqref{eq:DWmassMS})
where the negative mass region forms a strip with width $L/2$ (strip on KB; $P\pm, A\pm$).
Another setup is to put a circular domain-wall with radius $3L/8$ on an 
$\RPt$ (given by Eq.~\eqref{eq:DWmassMS2}), 
where the negative mass region is assigned to out of the domain-wall ($\RPt -\mathrm{disk};\pm\mp$).
As Fig.~\ref{fig:batas_MS} shows, our numerical results all converge 
to the expected values $\pm 1$ in the continuum theory where two distinct $\pinm$ structures are known.
Note that the above two setups have completely different shapes although 
they share the same topology.
Our observation supports that our formulation of the ABK invariant
$\beta^\mathrm{latt}$ is a topological invariant, being insensitive to details of the lattice geometries.

\begin{figure}[htbp]
  \centering
  \includegraphics[width=0.8\linewidth]{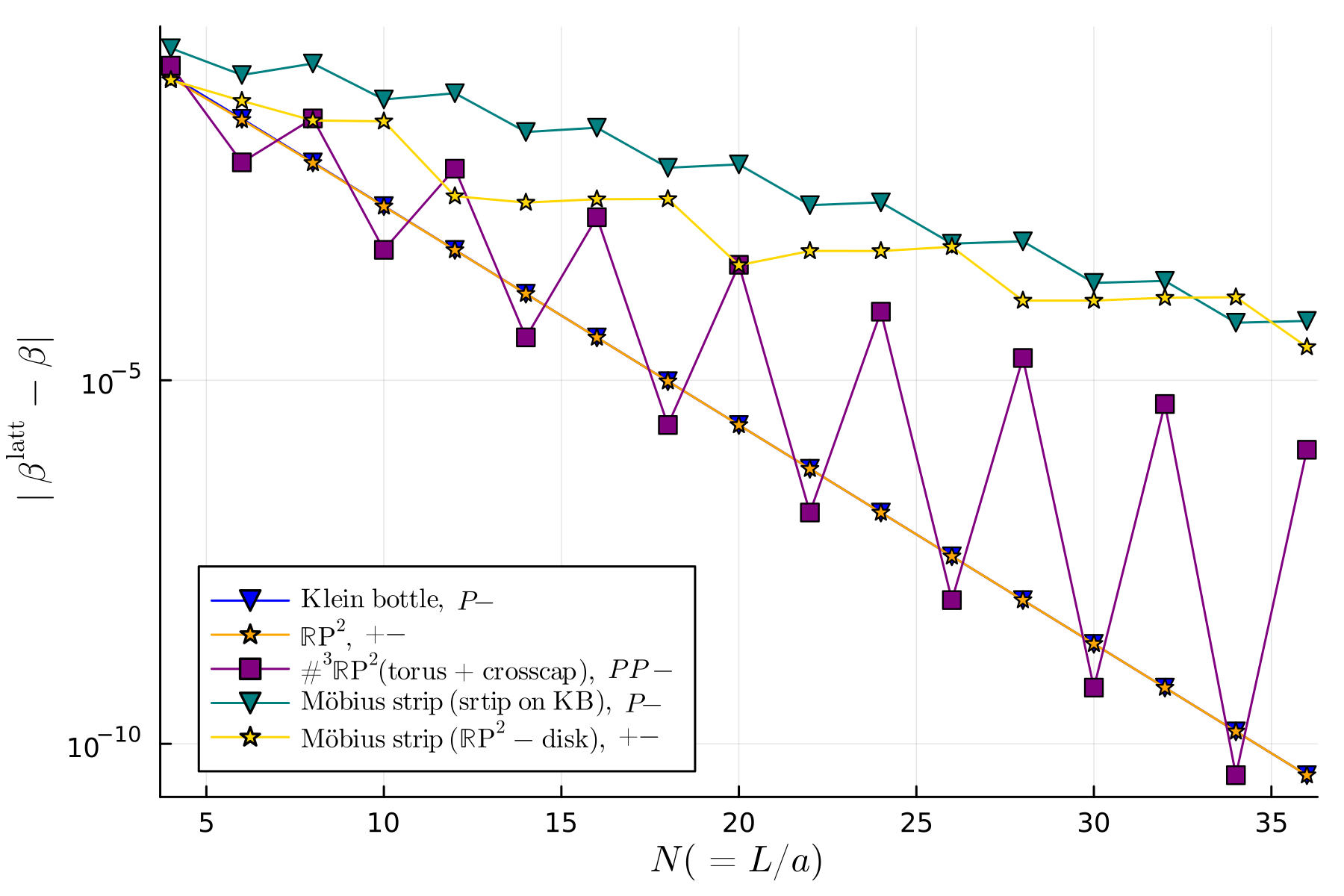} 
  \caption{Deviation of the lattice ABK invariant from the continuum value $|\beta^\mathrm{latt}-\beta|$ 
is plotted as a function of $N=1/a$ for various lattice setups.}
  \label{fig:batas_err}
\end{figure}

Figure~\ref{fig:batas_err} shows deviation of the lattice ABK invariants from the corresponding continuum theory values
in a logarithmic plot.
Although some of them show oscillating behaviors, they all show an exponential decay to zero,
which confirms that the lattice discretization systematics in our formulation of $\beta^\mathrm{latt}$ is well under control.

\begin{figure}[htbp]
  \centering
  \includegraphics[width=0.8\linewidth]{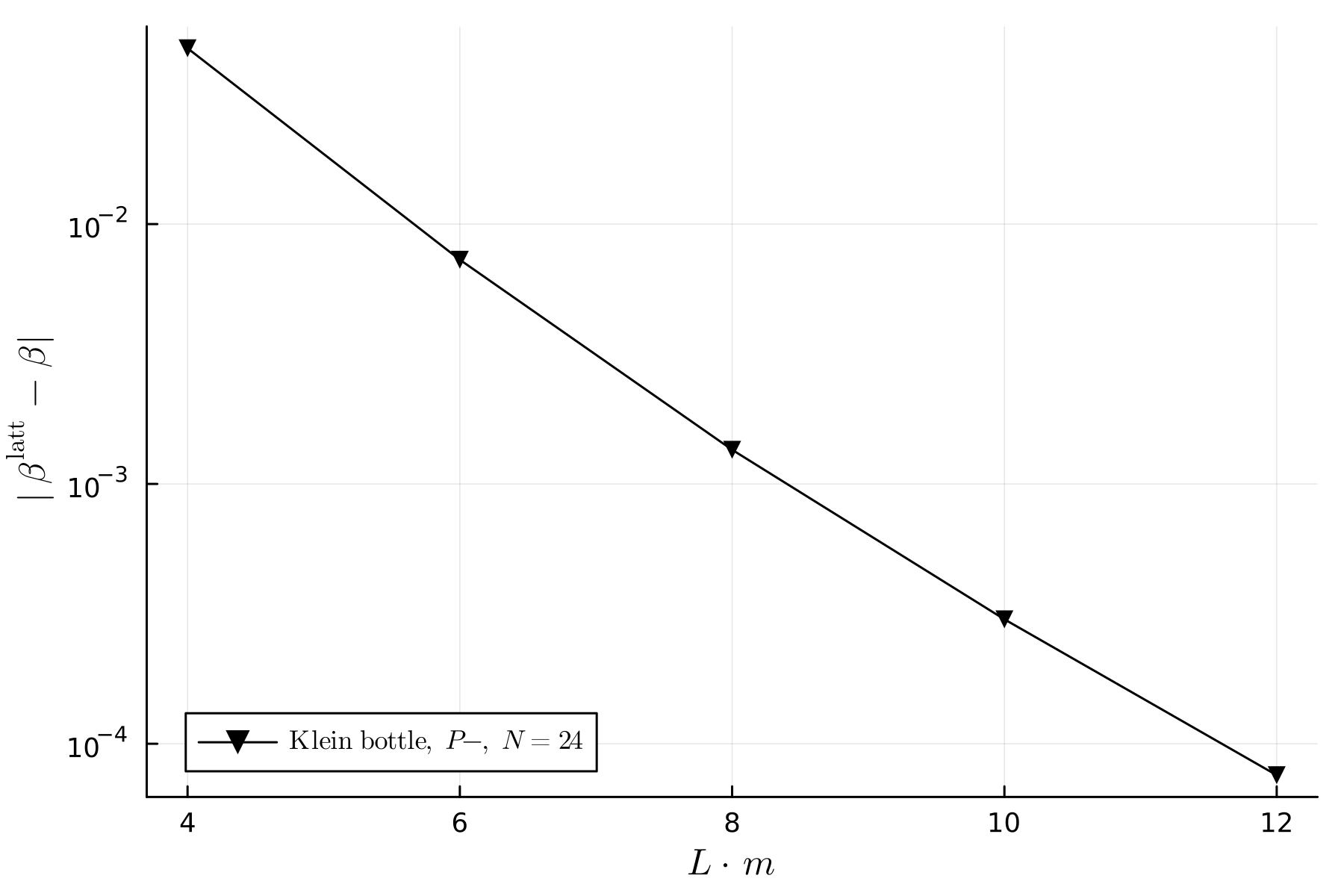} 
  \caption{Deviation of the lattice ABK invariant from the continuum value $|\beta^\mathrm{latt}-\beta|$ for the Klein bottle case is plotted as a function of $mL$ at a fixed value of $N=1/a=24$.}
  \label{fig:physmass_change}
\end{figure}

In order to see the finite fermion mass corrections, 
we also investigate the mass dependence of $\beta^\mathrm{latt}$ with a fixed $a=1/24$.
As Fig.~\ref{fig:physmass_change} shows, the deviation of $\beta^\mathrm{latt}$ from
its continuum value in the case of the Klein bottle exponentially decreases 
as a function of $m L$.

\clearpage
\section{Summary and discussion}
\label{sec:summary}
We have proposed a lattice formulation of the ABK invariant $\beta^\mathrm{latt}$ 
using the Pfaffian of the Wilson Dirac operator
that is a finite product of complex numbers. 
We have successfully equipped various nontrivial topology with nontrivial $\pinm$ structures of Majorana fermions by introducing 
twisted boundary conditions which reverse the orientation, 
as well as introducing cross-caps with twisted link variables.
Moreover, our formulation with the domain-wall mass term enables us 
to compute the ABK invariant on open manifolds without introducing the nonlocal APS boundary conditions.

On two-dimensional flat lattices, we have numerically verified 
that $\beta^\mathrm{latt}$ on a torus, Klein bottle, $\RPt$, ${\#}^3\RPt$, and two types of M\"obius strips, 
is quantized to the $\ZZ_8$ values all consistent with those in continuum theory.
It is remarkable that the convergence to the continuum limit is exponential with respect to
the lattice spacing $a$, rather than linear.
The underlying reason deserves further investigations.
It is also notable that $\RPt$ and ${\#}^3\RPt$ have nontrivial curvatures in continuum theory.
It seems that the singular plaquettes at the corners of our lattice 
have managed to represent the non-flat nature of these manifolds.
Moreover, the two different shapes of the M\"obius strips have shared the same $\beta^\mathrm{latt}$,
which supports that our formulation correctly captures the topology 
of the target manifold in the continuum limit, which is insensitive to details of the lattice geometry.

From the viewpoint of the lattice Dirac operator spectrum, 
it is rather nontrivial that the total complex phase of the lattice eigenvalues 
(including those of doublers) converges to a quantized value in $\ZZ_8$.
For two cases which have the translational symmetries, 
we have analytically identified 
the Fourier modes which contribute to $\beta^\mathrm{latt}$.
On a torus, the value is essentially determined by the zero modes,
while on a Klein bottle, $p_x=0$ but all $p_y\neq 0$ modes contribute.
For the latter case, we have explicitly shown that the summation of the complex phases
forms a good approximation of an integral which ends up with $2\pi$.

From the above observations, our lattice formulation gives a
well-defined regularization of the ABK invariant 
with the discretization systematics well under control.
We would like to emphasize that our formulation employs the
Wilson Dirac operator, without relying on any lattice chiral symmetry and the associated index theorem.

Finally we would like to comment on possible applications of this work.
The ABK invariant is a special version of the $\eta$-invariant, 
and our lattice formulation in this work would naturally apply to 
general types of the $\eta$-invariants, such as $\ZZ_{16}$-valued 
invariant that appears in the four dimensional Majorana fermion theory with $\pinp$ structures,
although there may be technical difficulties due to numerical costs.
In this work, we have focused on free Majorana fermion systems.
It would be interesting to investigate what role $\beta^\mathrm{latt}$ plays
in the interacting theories, in particular, when the number of fermions is eight, with which the anomaly cancels.
When there is a domain-wall, like the M\"obius strips we have studied in this work,
there would be a nontrivial bulk-edge anomaly cancellation or anomaly inflow.
Our lattice formulation would offer a nonperturbative tool to explore this issue.

\acknowledgments

We thank Mikio Furuta, Naoto Kan, Shinichiroh Matsuo and Kazuya Yonekura for useful discussion.
This work was
supported in part by JSPS KAKENHI Grants No.
JP21K03574, JP23K03387, JP23K22490 and JP25K07283, Japan.
This work was also supported in part by JST SPRING Grants No.
JPMJPS2138, Japan.

\appendix

\section{Fermions on a Klein bottle}\label{app:KB_eigenvectores}

A Klein bottle of size $L_x \times L_y$ can be constructed as a quotient of its double cover, a torus of size $L_x \times 2 L_y$ by an equivalence relation $(x,y+L_y)\sim(L_x-x,y)$.
The Majorana fermion field $\psi_{KB}(x,y)$ defined on the Klein bottle can be obtained by projecting the field $\psi_T(x,y)$ defined on the torus to the quotient.
For simplicity we consider this program in continuum theory but it is straightforward to
extend the discussion to the lattice gauge theory.

Let us start with imposing the boundary conditions on $\psi_T(x,y)$.
We take the antiperiodic condition in the $y$ direction\footnote{
With periodic boundary condition in the $y$ direction, it is difficult to maintain $\pinm$ structures.},
while both periodic($P$) and anti-periodic$(A)$ conditions are allowed in the $x$ direction:
\begin{align}
 \psi_T(x,y+2L_y) = -\psi_T(x,y),\;\;\; \psi_T(x+L_x,y) = s_x \psi_T(x,y).
\end{align}
where the sign $s_x$ is $+$ for $P$ and $-$ for $A$.

Next we introduce a glide reflection transformation by
\begin{equation}
  G_x \psi_T(x,y) = \gamma^1 \bar{\gamma}\psi_{T}\left(L_x-x,y -L_y\right),
\end{equation}
which satisfies $G_x^2=1$ and
the free Majorana fermion action is invariant under $G_x$, since $G_x$ commutes with $(D+m)$.
In fact, we can define the fermion fields on the Klein bottle 
by the projection of $\psi_T(x,y)$ onto the $G_x=s(=\pm 1)$ subspace:
\begin{equation}
 \psi^{s}_{KB}(x,y) := \frac{G_x+s}{2} \psi_T(x,y).
\end{equation}

It is not difficult to confirm that $\psi^{s}_{KB}(x,y)$ satisfy the
required boundary conditions on the Klein bottle,
\begin{align}
  &\psi^{s}_{KB}\left(x,L_y\right) = s R_x\psi^{s}_{KB}(x,0) = s \gamma^1 \bar{\gamma}\psi^{s}_{KB}\left(L_x-x,0\right), \label{eq:BCKByapp}\\
  &\psi^{s}_{KB}\left(L_x,y\right) = s_x \psi^{s}_{KB}\left(0,y\right),
\end{align}
and the fundamental region for the quotient by $G_x$ can be taken a half of the torus in the range $0\leq y <L_y$.
In particular, if $\psi_T(x,y)$ is an eigenfunction of $(D+m)$ (or more general operators which commute with $G_{x}$) then the projected $\psi^{s}_{KB}\left(x,y\right)$ is also an eigenfunction of $(D+m)$ with the same eigenvalue.

It is then not difficult to perform
the Fourier transformation of $\psi^{s}_{KB}(x,y)$ from that of $\psi_T(x,y)$
to obtain the eigenvalue decomposition of $\Pf [C(D+m)]$ as explicitly presented in Sec.~\ref{sec:KBFourier}.

\bibliographystyle{utphys.bst}
\bibliography{main_bibs}

\end{document}